\documentclass[preprint,sort&compress,11pt]{elsarticle}

\usepackage{graphicx}
\usepackage{amsmath}

\usepackage{natbib}
\usepackage{amssymb}

\usepackage{amsthm}

\usepackage{lineno}
\usepackage{subfig}
\usepackage{enumerate}
\usepackage{fullpage}
\usepackage{algorithm}
\usepackage{algpseudocode}
\usepackage{xcolor}
\usepackage[colorinlistoftodos]{todonotes}

\usepackage{hyperref}
\biboptions{sort,compress} 

\usepackage{xcolor}

\newcommand*\patchAmsMathEnvironmentForLineno[1]{%
  \expandafter\let\csname old#1\expandafter\endcsname\csname #1\endcsname
  \expandafter\let\csname oldend#1\expandafter\endcsname\csname end#1\endcsname
  \renewenvironment{#1}%
     {\linenomath\csname old#1\endcsname}%
     {\csname oldend#1\endcsname\endlinenomath}}% 
\newcommand*\patchBothAmsMathEnvironmentsForLineno[1]{%
  \patchAmsMathEnvironmentForLineno{#1}%
  \patchAmsMathEnvironmentForLineno{#1*}}%
\AtBeginDocument{%
\patchBothAmsMathEnvironmentsForLineno{equation}%
\patchBothAmsMathEnvironmentsForLineno{align}%
\patchBothAmsMathEnvironmentsForLineno{flalign}%
\patchBothAmsMathEnvironmentsForLineno{alignat}%
\patchBothAmsMathEnvironmentsForLineno{gather}%
\patchBothAmsMathEnvironmentsForLineno{multline}%
}

\usepackage{color,soul}
\usepackage{enumitem}
\usepackage{mathtools}
\usepackage{booktabs}

\definecolor{lightblue}{rgb}{.90,.95,1}
\definecolor{darkgreen}{rgb}{0,.5,0.5}

\definecolor{lightgreen}{rgb}{.90,1,0.90}

\newcommand{\bstau}{\boldsymbol{\tau}}

\usepackage{changes}
\definechangesauthor[name={Reviewer 1}, color = red]{Editor}

\usepackage{array}
\newcolumntype{P}[1]{>{\centering\arraybackslash}m{#1}}

\graphicspath{ {./figs/} }

\journal{Flow Turbulence Combust}

\begin{document}

\begin{frontmatter}

%\tableofcontents
%\listoftodos

\clearpage


  \title{A~Priori Assessment of Prediction Confidence for Data-Driven Turbulence Modeling}



\author[vt]{Jin-Long Wu}
\ead{jinlong@vt.edu}
\author[vt]{Jian-Xun Wang}
% \ead{vtwjx@vt.edu}
\author[vt]{Heng Xiao\corref{corxh}}
\cortext[corxh]{Corresponding author. Tel: +1 540 231 0926}
\ead{hengxiao@vt.edu}
\author[snl]{Julia Ling}
% \ead{jling@sandia.gov}

\address[vt]{Department of Aerospace and Ocean Engineering, Virginia Tech, Blacksburg, VA 24060, USA}
\address[snl]{Thermal/Fluid Science and Engineering, Sandia National Laboratories, Livermore, California 94551, USA}

\begin{abstract}
  Although Reynolds-Averaged Navier--Stokes (RANS) equations are still the dominant tool for
  engineering design and analysis applications involving turbulent flows, standard RANS models are
  known to be unreliable in many flows of engineering relevance, including flows with separation,
  strong pressure gradients or mean flow curvature. With increasing amounts of 3-dimensional
  experimental data and high fidelity simulation data from Large Eddy Simulation (LES) and Direct
  Numerical Simulation (DNS), data-driven turbulence modeling has become a promising approach to
  increase the predictive capability of RANS simulations. However, the prediction performance of data-driven models inevitably depends on the choices of training flows. This work aims to identify a quantitative measure for \textit{a priori} estimation of prediction confidence in data-driven turbulence modeling. This measure represents the distance in feature space between the training flows and the flow to be predicted. Specifically, the Mahalanobis distance and the kernel density estimation (KDE)
  technique are used as metrics to quantify the distance between flow data sets in feature space. To
  examine the relationship between these two extrapolation metrics and the machine learning model
  prediction performance, the flow over periodic hills at $Re=10595$ is used as test set and seven
  flows with different configurations are individually used as training sets. The results show that
  the prediction error of the Reynolds stress anisotropy is positively correlated with Mahalanobis
  distance and KDE distance, demonstrating that both extrapolation metrics can be used to
  estimate the prediction confidence \textit{a priori}. A quantitative comparison using correlation
  coefficients shows that the Mahalanobis distance is less accurate in estimating the prediction
  confidence than KDE distance. The extrapolation metrics introduced in this work and the
  corresponding analysis provide an approach to aid in the choice of data source and to assess the
  prediction performance for data-driven turbulence modeling.

\end{abstract}

\end{frontmatter}

%\linenumbers

\section{Introduction}
\label{sec:intro}

Even with the rapid growth of available computational resources, numerical models based on Reynolds-Averaged 
Navier--Stokes (RANS) equations are still the dominant tool for engineering design and
analysis applications involving turbulent flows. However, the development of turbulence models has
stagnated--the most widely used general-purpose turbulence models (e.g., $k$-$\varepsilon$
models, $k$-$\omega$ models, and the S--A model) were all developed decades ago.  These models are
known to be unreliable in many flows of engineering relevance, including flows with
three-dimensional structures, swirl, pressure gradients, or curvature~\cite{Craft}.  This lack
of accuracy in complex flows has diminished the utility of RANS as a predictive simulation tool for
use in engineering design, analysis, optimization, and reliability assessments.

Recently, data-driven turbulence modeling has emerged as a promising alternative to traditional
modeling approaches.  While data-driven methods come in many formulations and with different
assumptions, the basic idea is that a model or correction term is determined based on data.  In the
context of turbulence, this data can come from either experiments or high fidelity simulations such
as Direct Numerical Simulations (DNS) or well-resolved Large Eddy Simulations (LES).
Koumoutsakos~\cite{Milano} trained neural networks on channel flow DNS data and applied them
to the modeling of near-wall turbulence structures.  Tracey et al.~\cite{Tracey2013} used neural
networks to predict the Reynolds stress anisotropy and source terms for transport equations of
turbulence quantities (e.g., $\tilde{\nu}_t$ for the S--A model and $\omega$ for $k$-$\omega$
models).  Duraisamy et al.~\cite{Tracey2013, Duraisamy2015} have used Gaussian processes to
predict the turbulence intermittency and correction terms for the turbulence transport equations.
Ling and Templeton~\cite{ling2015evaluation} trained random forest classifiers to predict when RANS assumptions would
fail.  Ling et al.~\cite{Ling2016,ling2016reynolds} further used random forest regressors and neural networks to predict the Reynolds
stress anisotropy.  Wang et al.~\cite{Wang2016} have recently investigated the use of random forests
to predict the discrepancies of RANS modeled Reynolds stresses in separated flows.  These studies
show the significant and growing interest in applying data-driven machine learning techniques to
turbulence modeling.

However, Tracey et al.~\cite{Tracey2013}, Ling and
Templeton~\cite{ling2015evaluation}, and Wang et al.~\cite{Wang2016} all reported that their data-driven
closures had diminished performance on flows that were significantly different from the ones on
which they were trained. Such findings underline the importance of properly choosing the training flows. For instance, if a machine learning model for
 the eddy viscosity is trained on a database of attached boundary layer
flows, then it would not be surprising if the model had poor performance when making predictions on
a flow with separation. In the general context of data-driven modeling, the
\textit{training set} is the set of data to which the model is fit or calibrated. The \textit{test
  set} is the set of data on which the model makes predictions. It is expected that the prediction performance will not be satisfactory if the test set is significantly different from the training sets.

A key question, then, is how to determine whether the test flow is ``different'' from the training
flows.  Test flow that might appear very different could, in fact, be well-supported in the training
set.  For example, a model trained on flow around an airfoil might perform well on a channel flow
because it will have encountered developing boundary layers in its training set.  Conversely, a
model trained on attached flow around an airfoil might perform very poorly on stalled flow around an
airfoil, even though the flow configurations appear quite similar.  Because many of these data-driven
models are formulated such that their inputs are the local flow variables, it is the local flow
regimes that must be well-supported in the training set, not any specific global geometry.  The
degree to which the test flow is well-supported by training data will in large part determine the
reliability of the model closure.  In deploying these data-driven models, then, it will be crucial
to have efficient metrics of determining if a test flow is an extrapolation from the training set.

Ling and Templeton~\cite{ling2015evaluation} presented one promising option for quantifying model
extrapolation.  They used a statistical metric called the Mahalanobis distance to calculate the
probability that a test point was drawn from the training distribution, and showed that their
machine learning model error was significantly higher on test points with high Mahalanobis
distances.  However, they did not carry out a thorough investigation of the correlation between the
Mahalanobis distance and machine learning model accuracy for different test sets and training sets.
Because the Mahalanobis distance assumes that the training data have a multivariate Gaussian
distribution, it is not clear how generally applicable this metric will be for different flow cases.
Therefore, this paper will investigate two different statistical metrics of extrapolation, the
Mahalanobis distance and the distance based on Kernel Density Estimation (KDE), for a variety of
training sets.  These extrapolation metrics will be analyzed and compared in the context of a machine learning framework for the prediction of Reynolds stresses developed by Wang
et al.~\cite{Wang2016}. The ultimate goal of this work is to provide quantitative metrics to assess the prediction confidence \textit{a priori}, in order to guide the choice of training set when applying data-driven turbulence modeling.

The rest of the paper is organized as follows. Section~\ref{sec:ml-methodology}
will describe the machine learning methodology of Wang et al.~\cite{Wang2016} and the database of
flows used for training and testing.  Section~\ref{sec:extrapolation} will describe different
extrapolation metrics and their relative advantages and disadvantages.  Section~\ref{sec:results} will
analyze the performance of these extrapolation metrics in predicting the regions of high uncertainty
in the machine learning models, and Section~\ref{sec:conclusion} will present conclusions and ideas for
next steps.

\section{Machine Learning Methodology}
\label{sec:ml-methodology} 

In this section we summarize the Physics-Informed Machine Learning (PIML) framework proposed by Wang
et al.~\cite{Wang2016}, which will be used to evaluate the efficacy of the proposed
extrapolation metrics.  However, note that extrapolation metrics proposed in this work are not
specific to the PIML framework.  Particular emphasis is placed on applications where the feature
space input has high dimensions (10 to 100 features), which is typical in computational mechanics
problems and in other complex physical systems (see e.g., \cite{ling2016jcp}).

The overarching goal of the work of Wang et al.~\cite{Wang2016} is a physics-informed machine
learning framework for turbulence modeling. The problem can be formulated as
follows: given high-fidelity data for the Reynolds stresses from DNS or well-resolved LES on a database of flows, predict the Reynolds stresses on a new flow for which only RANS data are available. The
flows used to fit the machine learning model are referred to as \emph{training flows},
while the flow for which the model is evaluated is referred to as the \emph{test flow}.  It is assumed that the
training flows and the test flow have similar flow physics.

Wang et al.~\cite{Wang2016} trained machine learning models to predict a corrector for the Reynolds stress tensor. The procedure is summarized as follows:
 \begin{enumerate}
 \item Perform baseline RANS simulations on both the training flows and the test flow.
 \item Compute the feature vector $\mathbf{q}(\mathbf{x})$ based on the RANS state variables.
 \item Compute the discrepancies field $\Delta \tau_\alpha(\mathbf{x})$ of the RANS modeled Reynolds
   stresses for the training flows based on the high-fidelity data, where $\Delta \tau_\alpha =
   \tau^{RANS}_\alpha - \tau^{truth}_\alpha$.
 \item Construct regression functions $ f_\alpha: \mathbf{q} \mapsto \Delta \tau_\alpha$ for the
   discrepancies based on the training data prepared in Step 3. These regression functions were constructed using random forest regressors~\cite{liaw2002classification}.
 \item Compute the Reynolds stresses discrepancies for the test set by evaluating the
   regression functions. The Reynolds stresses can subsequently be obtained by correcting the
   baseline RANS predictions with the evaluated discrepancies.
 \end{enumerate}

 In machine learning terminology the discrepancies $\Delta \tau_i$ here are referred to as
 \emph{responses}, the feature vector $\mathbf{q}$ as the \emph{input}, and the mappings $ f_\alpha: \mathbf{q}
 \mapsto \Delta \tau_\alpha$ as \emph{regression functions}.

 There are three essential components in the physics-based machine learning framework outlined
 above: (1) identification of mean flow features as input, (2) representation of Reynolds stresses
 as responses, and (3) construction of regression functions from training data.  The three
 components of the framework are presented below.  The reader is referred to~\cite{Wang2016} for
 further details.

\subsection{Identification of Mean Flow Features as Regression Input}

Ten features based on the RANS computed mean flow fields (velocity $U_i$ and pressure $P$) are
identified as inputs to the regression function, which is consistent with the mean flow features listed in the work by Wang et al.~\cite{Wang2016}. Most of these features are adopted from the work by Ling and
Templeton~\cite{ling2015evaluation}, with an additional feature of mean streamline curvature. It is because that RANS models tend to be less reliable at the regions with large streamline curvature, thus including it as a mean flow feature helps in detecting those regions. Turbulence intensity is another mean flow feature in the work by Wang et al.~\cite{Wang2016}, since it is an important feature in describing turbulence. Similarly, the turbulence time scale is also chosen as a feature to better describe the turbulence. Wang et al.~\cite{Wang2016} also chose the wall-distance based Reynolds number as a mean flow feature, since the presence of wall dampens the wall normal fluctuation and thus has a important impact upon the Reynolds stress anisotropy. The pressure gradient along streamline is also used as a mean flow feature, even though a uniform acceleration of an incompressible flow does not alter the turbulence. The main reason is that Reynolds stress discrepancy between the RANS simulation and DNS/LES simulation is to be predicted, and this discrepancy is influenced by $dp/dx_i$. For example, the RANS simulation is usually more reliable for the equilibrium boundary layer, but is less reliable for the boundary layer with strong acceleration. Therefore, larger Reynolds stress discrepancy can be expected for the boundary layer with strong acceleration, and the inclusion of $dp/dx_i$ helps in distinguishing such scenario. Most of the mean flow features are normalized within the range between -1 to 1, except for the wall-distance based on Reynolds number that lies within the range between 0 to 2. With the same length of range, the Euclidean distance along different features is
comparable to each other. The normalization also facilitate machine learning and is a common
practice there.

These mean flow features are adopted from the data-driven turbulence modeling
  framework~\cite{Wang2016} to ensure consistency. A detailed list of the ten mean flow features used in this work can be found in the work by Wang et al.~\cite{Wang2016}. However, it should be noted that the current work focuses on the investigation of a priori assessing the closeness of flow features between the training flows and the test flow, and the \textit{a priori} confidence assessment metrics in this work is directly applicable to other choice of mean flow features. All these features are independent under rotation, translation or reflection of the coordinate system. However, some of them are not Galilean invariant, e.g., the normalization factor $U_iU_i$ and the streamline curvature $D \bold{\Gamma}/ Ds$ that are dependent on the velocity of a moving reference frame. Therefore, the authors recommend the use of fixed coordinate systems for both the training and prediction flows.

\subsection{Representation of Reynolds Stress Discrepancy as Regression Response} 

The discrepancies of RANS predicted Reynolds stresses, or more precisely the
magnitude, shape and orientation thereof, are
identified as responses of the regression functions.  It has been shown that these discrepancies are
likely to be universal among flows of the same characteristics, and thus the regression function
constructed based on them can be extrapolated to new flows~\cite{Wang2016}. To obtain the
components, the Reynolds stress tensor is decomposed as follows~\cite{emory2011modeling,xiao-mfu}:
\begin{equation}
  \label{eq:tau-decomp}
  \boldsymbol{\tau} = 2 k \left( \frac{1}{3} \mathbf{I} +  \mathbf{A} \right)
  = 2 k \left( \frac{1}{3} \mathbf{I} + \mathbf{V} \Lambda \mathbf{V}^T \right)
\end{equation} 	
where $k$ is the turbulent kinetic energy, which indicates the magnitude of $\bstau$; $\mathbf{I}$
is the second order identity tensor; $\mathbf{A}$ is the anisotropy tensor; $\mathbf{V} =
[\mathbf{v}_1, \mathbf{v}_2, \mathbf{v}_3]$ and $\Lambda = \textrm{diag}[\lambda_1, \lambda_2,
\lambda_3]$ with $\lambda_1+\lambda_2+\lambda_3=0$ are the orthonormal eigenvectors and eigenvalues
of $\mathbf{A}$, respectively, indicating its shape and orientation.

\begin{figure}[!htbp]
  \centering
   \includegraphics[width=0.5\textwidth]{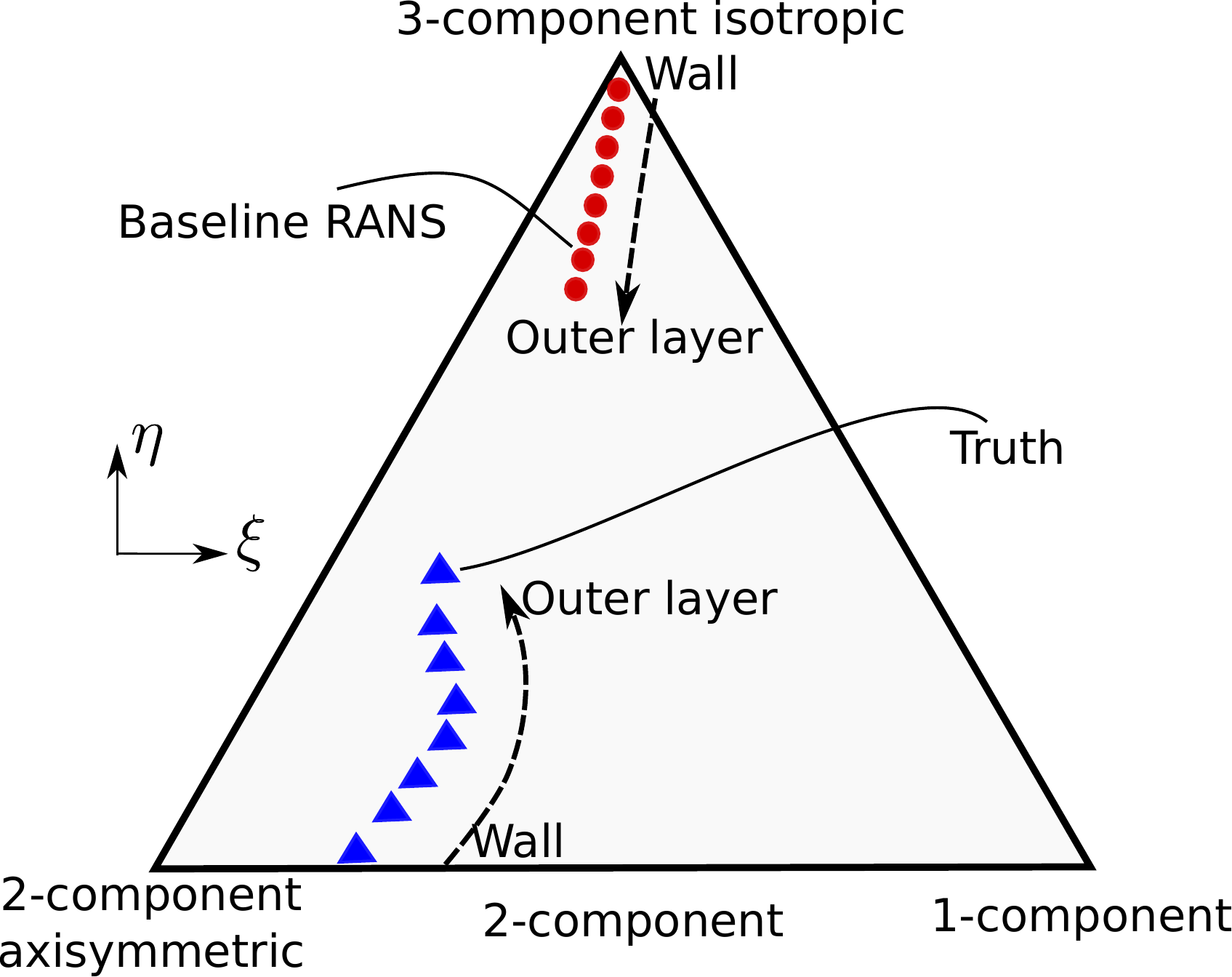}
   \caption{The Barycentric triangle that encloses all physically realizable states of the Reynolds
     stresses~\cite{banerjee2007presentation,emory2013modeling}. The position within the Barycentric
     triangle represents the anisotropy state of the Reynolds stress. The three corners represent
     the limiting states.}
  \label{fig:bary}
  %\todo[inline]{Replace the figure with the one explaining the Barycentric triangle.}
\end{figure}

In the Barycentric triangle shown schematically in Fig.~\ref{fig:bary}, the eigenvalues $\lambda_1$,
$\lambda_2$, and $\lambda_3$ are mapped to the Barycentric coordinates as
follows~\cite{banerjee2007presentation}:
\begin{subequations}
	\label{eq:lambda2c}
        \begin{align}
          C_1 & = \lambda_1 - \lambda_2 \\
          C_2 & = 2(\lambda_2 - \lambda_3) \\
          C_3 & = 3\lambda_3 + 1 \  ,
        \end{align}  
\end{subequations}
where $C_1 + C_2 + C_3 = 1$.  Placing the triangle in a Cartesian coordinate $\boldsymbol{\xi}
\equiv (\xi, \eta)$, the location of any point within the triangle is a convex combination of those
of the three vertices, i.e.,
\begin{equation}
\boldsymbol{\xi} = 	\boldsymbol{\xi}_{1c}C_1 + \boldsymbol{\xi}_{2c}C_2 +
\boldsymbol{\xi}_{3c}C_3 
\end{equation}	
where $\boldsymbol{\xi}_{1c}$, $\boldsymbol{\xi}_{2c}$, and $\boldsymbol{\xi}_{3c}$ denote coordinates of the
three vertices of the triangle. Consequently, the coordinate $\boldsymbol{\xi} \equiv (\xi, \eta)$
uniquely identifies the shape of the anisotropy tensor.

In this work, the discrepancies of the coordinate $\Delta \boldsymbol{\xi} \equiv (\Delta \xi, \Delta \eta)$ are chosen as the regression responses, where $\Delta \xi = \xi_{DNS}-\xi_{RANS}$ and $\Delta \eta = \eta_{DNS} - \eta_{RANS}$ represent the discrepancy between the RANS predicted Reynolds stress anisotropy and the DNS data.

\subsection{Random Forest for Building Regression Functions}

With the input (mean flow features $\mathbf{q}$) and responses (Reynolds stress discrepancies
$\Delta\tau_i$) identified above, an algorithm is needed to map from the input to the responses. In
this work, random forest regression is employed~\cite{breiman2001random}. Random forest regression
is an ensemble learning technique that aggregates predictions from a number of decision trees. In
decision tree learning, a tree-like model is built to predict the response variable by learning
simple decision rules from the training data.  While decision trees have the advantages of being
computationally efficient and amenable to interpretation, they tend to overfit the data, i.e., yield
models that reproduce the training data very well but predict poorly for unseen data. In random forest
regression, an ensemble of trees is built with bootstrap aggregation samples (i.e., sampling with
replacement) drawn from the training data~\cite{friedman2001elements}.  Moreover, only a subset of
randomly chosen features is used when constructing each split in each tree, which reduces the
correlation among the trees in the ensemble.
By aggregating a large number of trees obtained in this manner, random forests can achieve
significantly improved predictive performance and largely avoid overfitting. In
  addition, random forests can provide an relative importance score for each input feature by
  counting the times of splitting within the decision trees based on the given flow feature. These
  importance scores reflect the influence of the choice of flow features on the training-prediction
  performance. More detailed discussion of the feature importance can be found in the work by Wang
  et al.~\cite{Wang2016}. Random forest regression is a widely used regression
  method in machine learning community. Compared to the neural network, the random forest is less
  prone to overfitting and is thus more robust as pointed out by Breiman~\cite{breiman2001random}.

\section{Extrapolation Metrics}
\label{sec:extrapolation}

\subsection{Motivation with Machine Learning Based Predictive Turbulence Modeling}
\begin{figure}[!htbp]
  \centering
  \subfloat[Periodic hills, $Re=10595$]{\includegraphics[width=0.45\textwidth]{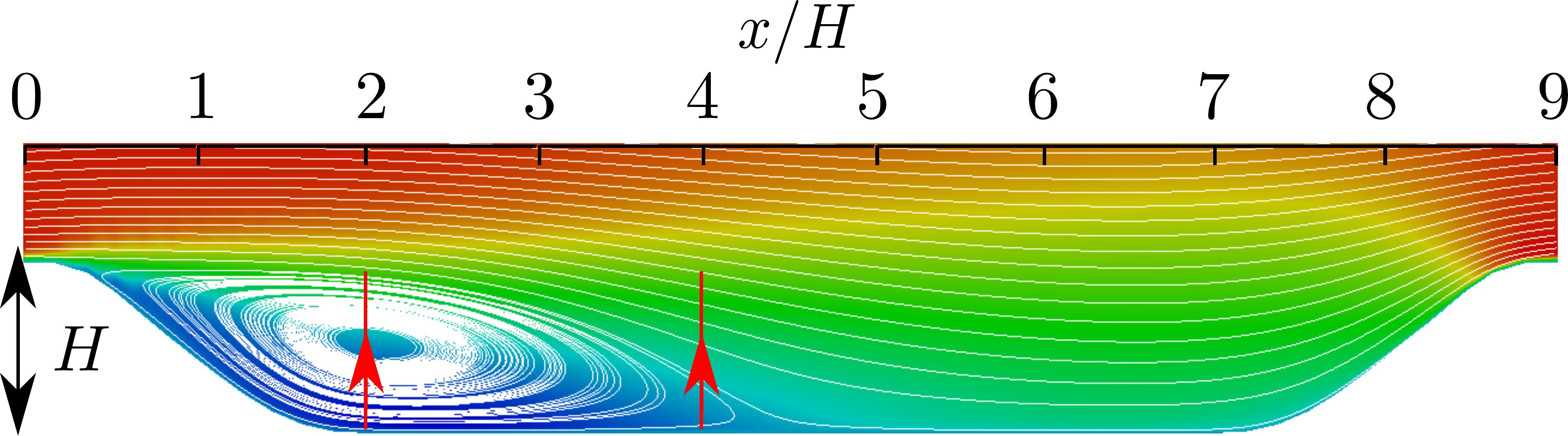}}\hspace{0.5em}\\
  \subfloat[Periodic hills, $Re=5600$]{\includegraphics[width=0.45\textwidth]{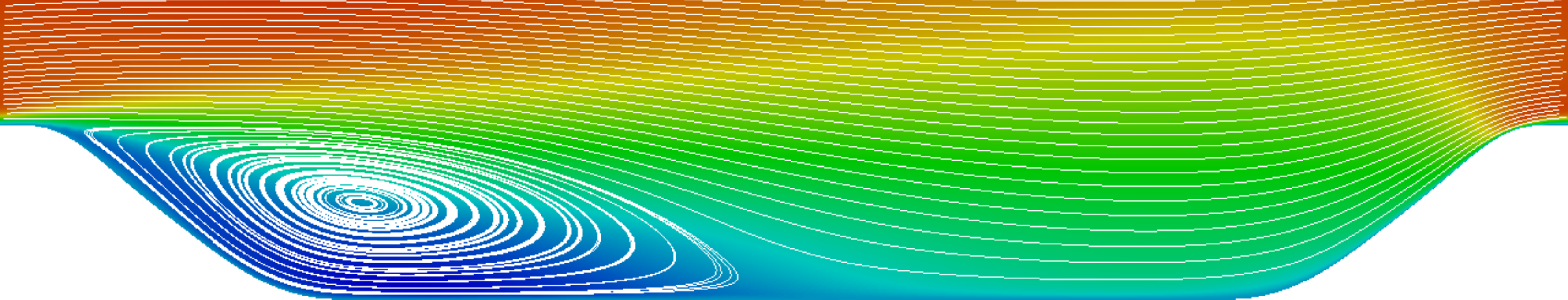}}\hspace{0.5em}  
  \subfloat[Curved backward step, $Re=13200$]{\includegraphics[width=0.35\textwidth]{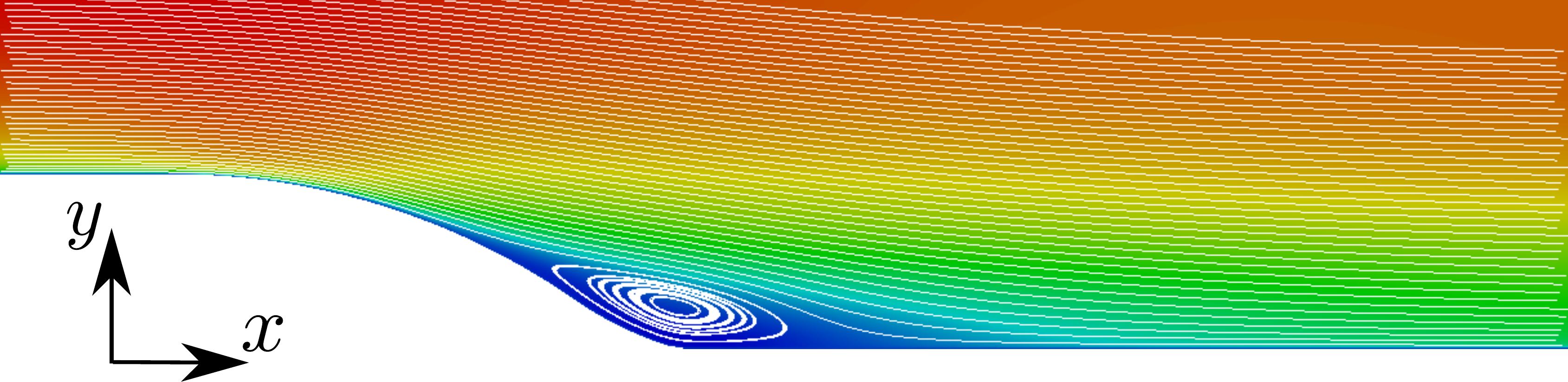}}\\
  \subfloat[Converging--diverging channel, $Re=11700$]{\includegraphics[width=0.48\textwidth, height=0.08\textwidth]{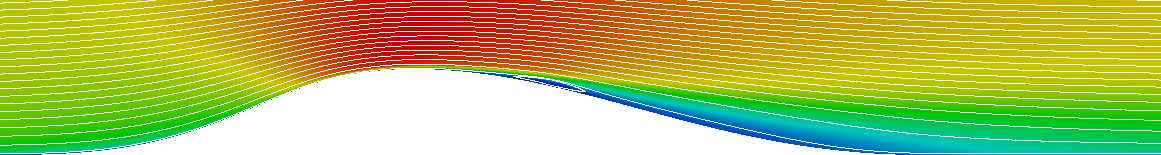}}\hspace{0.5em}
  \subfloat[Backward facing step $Re=4700$]{\includegraphics[width=0.32\textwidth, height=0.08\textwidth]{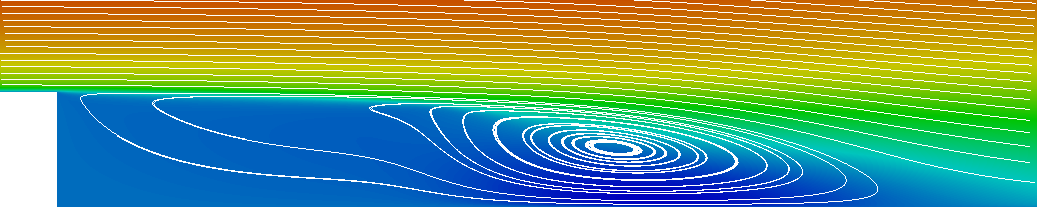}}\\  
  \subfloat[Wavy channel, $Re=360$]{\includegraphics[width=0.7\textwidth, height=0.1\textwidth]{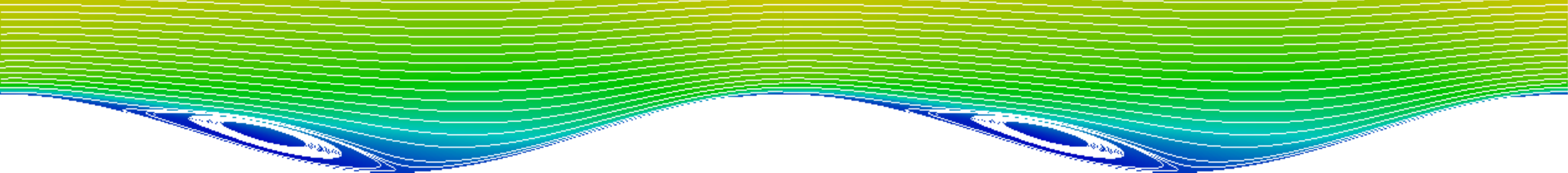}}
  \caption{The mean velocity field and separation bubble of the test set and different training sets. The test set is (a) the flow over periodic hills at $Re=10595$. The training
    sets include (b) the flow over periodic hills at $Re=5600$, $Re=2800$ and $Re=1400$ (only the flow at $Re=5600$ is shown here for simplicity), (c) the flow over curved backward
    facing step at $Re =13200$, (d) the flow in a converging-diverging channel at $Re =
    11300$, (e) the flow over a backward facing step at $Re = 4900$, and (f) the flow in a
    channel with wavy bottom wall at $Re = 360$. The test set is the flow over periodic hills at
    $Re=10595$. All the flow fields illustrated in this figure are obtained based on RANS simulations, in which Launder-Sharma $k$-$\epsilon$ model~\citep{launder1974application} is used. The lines with arrows in panel (a) indicate the profile locations for the
    anisotropy presented in Figs.~\ref{fig:bary2} and~\ref{fig:bary4}.}
\label{fig:case-intro}
\end{figure}

The machine learning framework as summarized in Section~\ref{sec:ml-methodology}
was used by Wang et al.~\cite{Wang2016} to predict Reynolds stresses in conjunction with standard
RANS models. The objective was to predict the Reynolds stress in the flow over period
hills at $Re=10595$.  Training flows were chosen from the NASA benchmark
database~\cite{nasa-web} including (1) the flow over periodic hills at $Re=1400$ (PH1400), 2800 (PH2800) and 5600 (PH5600)~\cite{breuer2009flow}, (2) the flow past a curved
backward facing step (CBFS13200) at $Re=13200$~\cite{bentaleb2012large}, (3) the flow in a converging-diverging channel (CDC11300) at Re=$11300$~\cite{laval2011direct},
(4) the flow past a backward facing step (BFS4900) at $Re=4900$~\cite{le1997direct}, and (5) the flow in a channel with wavy bottom wall (WC360) at $Re=360$~\cite{maass1996direct}. The geometries and the flow characteristics of these flows are
illustrated in Fig.~\ref{fig:case-intro}. The Reynolds numbers are defined based on the bulk
velocity $U_b$ at the narrowest cross-section in the flow and the height $H$ of the crest or
step.

In this study we conducted an extensive evaluation of the machine learning based prediction of
Reynolds stresses, in the context of extrapolation detection. In order to isolate the contribution
of the data from each flow to the predictive capability and to simplify the performance assessment,
each flow above is individually used for training as opposed to combining several flows.  In
general, our experiences suggest that better predictive performance is obtained when the training
flow and the prediction flow are more similar.  For example, since the test case is the flow over
periodic hills at $Re=10595$, the predictive performance is the best when the training case is the
flow in the same geometry but at a different Reynolds number $Re=5600$.  In contrast, the
performance is the least favorable when the backward step flow $Re=4900$ or the wavy channel flow
$Re=360$ is used as training flow.  Physical intuition suggests that the backward step and wavy
channel cases are the furthest from the prediction flow in feature space. This is because the
backward step has a sudden expansion (as opposed to the gradual expansion in the periodic hill
geometry) and the wavy channel flow has a Reynolds number ($Re=360$) that is drastically different
from that in the test case ($Re=10595$).  The other two flows (curved step and converging--diverging
channel) fall between the two extremes above because their geometries are qualitatively similar to
that of the test case, and their Reynolds numbers (13200 and 11700) are comparable to 10595 as well.
Hence, the predictive performances of different training cases agree well with our physical
understanding of degrees of similarity between the training and test flows.  However, when
data-driven models are used in practical flows, the similarity between the prediction flow and
various candidate sets of training flows is usually not clear. The performances of data-driven models depend on the similarity
  between the training flows and the test flow. Predicting the test flow with significantly
  different flow physics from the training flows is potentially catastrophic. To prevent these
  consequences that may stem from bad judgment from users of data-driven models, in this work we
  propose the extrapolation metrics that objectively quantifies the similarity between the training
  flows and the test flow. These extrapolation metrics can assess the prediction performance of the
  data-driven model \textit{a priori}, and they also have the potential to provide guidelines for
  selecting more suitable training flows to improve the prediction.

%\begin{figure}[!htbp]
%  \centering
%  \subfloat[periodic hills, $Re=5600$]{\includegraphics[width=0.4\textwidth]{Ucontour_PH_crop}}\hspace{0.5em}
%  \subfloat[curved backward step, $Re=13200$]{\includegraphics[width=0.4\textwidth]{Ucontour_CBFS}}\\
%  \subfloat[converging--diverging channel, $Re=11700$]{\includegraphics[width=0.4\textwidth]{Ucontour_CDC}}\hspace{0.5em}
%  \subfloat[backward facing step $Re=4700$]{\includegraphics[width=0.4\textwidth]{Ucontour_BFS}}\\  
%  \subfloat[wavy channel, $Re=360$]{\includegraphics[width=0.7\textwidth]{Ucontour_WC}}
%  \caption{The mean velocity field and separation bubble of different training sets. The training
%    sets include (a) the flow over periodic hills at $Re=5600$, (b) flow over curved backward
%    facing step at $Re =13200$, (c) flow in a converging-diverging channel at $Re =
%    11300$, (d) flow over a backward facing step at $Re = 4900$, and (e) flow in a wavy
%    channel at $Re = 360$. The prediction set is the flow over periodic hills at
%    $Re=10595$. The red line with arrow in panel (a) indicates the sampled lines along which the
%    anisotropy is shown in Fig.~\ref{fig:bary}.}
%\label{fig:case-intro}
%\end{figure}

\subsection{Extrapolation Metrics}
\label{sec:bad-metrics}

Before presenting the details of the two metrics investigated in this work, we examine a few
apparently attractive candidates of extrapolation metrics based on nearest neighbor distances and
marginal distributions. We discuss and illustrate why they are not suitable.

The idea behind extrapolation metrics is to determine the extrapolation distance between a given
test point or test set and the training data.  There are several different approaches for
quantifying this distance.  One metric would be the \emph{nearest neighbor distance}.  The nearest neighbor
distance is the Euclidean distance in feature space between the test point and the nearest point in
the training set.  Because this nearest neighbor distance is susceptible to noise, a common
variation is the $K^{th}$ nearest neighbor distance~\cite{Cover1967}, which is the Euclidean
distance to the $K^{th}$ nearest point in the training set, where $K$ is some pre-determined
integer.  Unfortunately, these methods are unwieldy--they require retaining the entire training database
 to compare against.  In turbulence simulations, even the mean flow data from a single
simulation can consume many gigabytes of memory, so transferring the training database to each user
of the machine learning model based on these training data sets is impractical.

\begin{figure}[!htbp]
  \centering
  \includegraphics[width=0.45\textwidth]{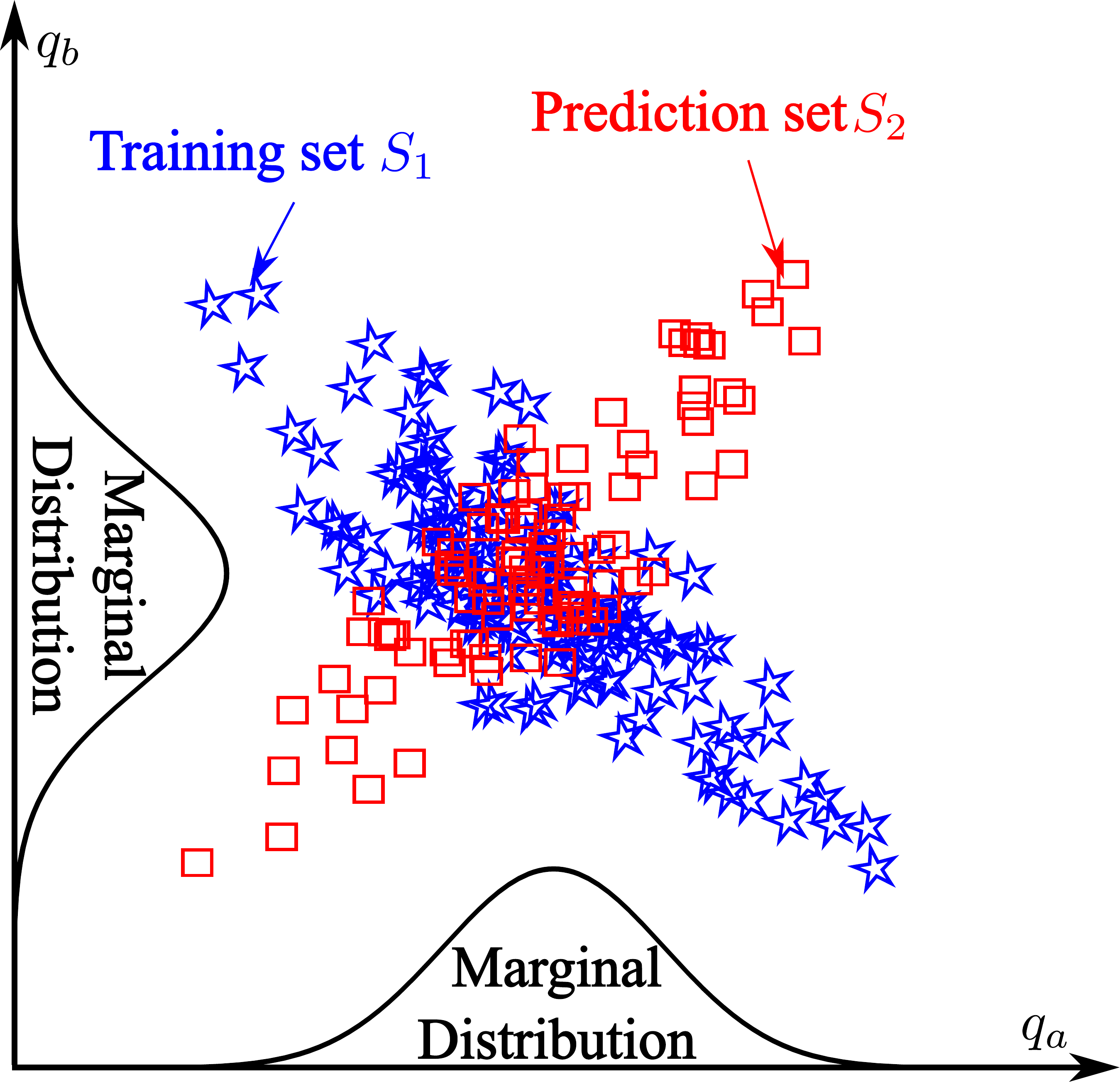} % png
  \caption{Illustration of two data sets that have identical marginal distributions but cover different regions in feature space.}
\label{fig:marginal}
\end{figure}

Another seemingly appealing yet equally unsuitable indicator of distance between two sets of points
is the \emph{marginal probability density functions}. This is illustrated in Fig.~\ref{fig:marginal}
for the simple example of a two-dimensional feature space. This figure shows two data sets $S_1$ and
$S_2$ have identical marginal densities (indicated by the bell-shaped curves on the two axes) but
cover distinctly different regions in the feature space. If $S_1$ is used for training to predict
the response of points in $S_2$, most of the evaluations would involve aggressive extrapolations,
and thus poor predictive performance would be expected. The situation will be even more pronounced
in higher-dimensional feature spaces.

After evaluating a number of alternatives, we identified the Mahalanobis distance and Kernel Density
Estimation (KDE) as two promising metrics for evaluating the predictive performance of
training--prediction methods. In both metrics, only the inputs of the training and test data are used
and information of the response is not needed. More importantly, only the statistical quantities
(e.g., mean, covariance, estimated probability density function of the training set) are used, and
the complete raw training data set is not needed. This characteristic leads to much lower memory
consumption, which is in contrast to the nearest neighbor distance methods.

The Mahalanobis distance~\cite{mahalanobis1936generalized} is an efficient method of
representing the extrapolation distance that does not rely on marginal distributions.  The
Mahalanobis distance is defined as the distance between a point $\tilde{\mathbf{q}}$ and the mean
of the training points $\boldsymbol{\mu}$, scaled by the covariance matrix $\Sigma$ of the
training points,
 \begin{equation}
   \label{eq:mdist}
   D = \sqrt{(\tilde{\mathbf{q}} - \boldsymbol{\mu})^T \Sigma^{-1} (\tilde{\mathbf{q}}
     -\boldsymbol{\mu})} 
 \end{equation}
 The basic idea of the Mahalanobis distance is that it measures the distance between a point and a
 distribution--in this case, the distribution of training points.  The larger the Mahalanobis
 distance, the greater the degree of extrapolation.  To determine the Mahalanobis distance of a test
 point, it is not necessary to compare against all of the training data--only the mean and
 covariance matrix of the training data must be saved, which is highly memory efficient.

 In this paper, the Mahalanobis distance has been normalized based on percentiles from the training
 set. The normalized Mahalanobis distance of a test point is given by $1 - \gamma_{dm}$, where
 $\gamma_{dm}$ is the fraction of training points with a larger raw Mahalanobis distance than the
 test point.  This normalization ensures that the distance lies in the range between 0 and 1, with a
 normalized distance of 0 indicating no extrapolation and a distance of 1 indicating very high
 extrapolation. With this normalization convention, larger raw Mahalanobis distance leads to larger normalized distance. This is because the raw Mahalanobis distance takes into account the scattering of the training flow points by incorporating the covariance matrix $\Sigma$. Therefore, larger raw distance means the distance from the test flow point to the mean of the training flow points is greater relative to the scattering of the training flow points. In other words, fewer training points have a greater raw distance than the raw distance of the test flow point, and thus a larger normalized Mahalanobis distance can be expected.

 An underlying assumption for the Mahalanobis distance is that the training set distribution is a
 multivariate Gaussian.  However, this may be a poor assumption in many turbulent flows, where
 multi-modal distributions may be common.  Therefore, the performance of the Mahalanobis distance as
 an extrapolation metric will be compared against Kernel Density Estimation (KDE)~\cite{Silverman}.
 KDE is a method of approximating the distribution of the training data.  The KDE at a point
 $\tilde{\mathbf{q}}$ from a distribution of training data $\mathbf{q}_{i}$ for $i=1, \cdots, n$ is:
\begin{eqnarray}
  \hat{f} = \frac{1}{n \sigma} \sum\limits_{i=1}^{n} K\left(\frac{ \tilde{\mathbf{q}} - {\mathbf{q}}_{i}}{\sigma}\right)
\end{eqnarray}
Often, a Gaussian kernel is used, and the bandwidth $\sigma$ is the standard deviation of the
Gaussian kernel. In this work, the bandwidth is determined based on Scott's rule~\cite{scott2015multivariate}.

Unlike the Mahalanobis distance, the Gaussian KDE distance does not assume that the training set has
a multivariate Gaussian distribution.  Instead, it has a much less stringent assumption--that the
training data distribution can be approximated as a sum of multivariate Gaussians.  KDE can be used
as an extrapolation metric by determining the probability that a given test point was drawn from the
training data. The KDE distance has been normalized by comparing the KDE probability estimation to a 
uniform distribution, as shown in Eq.~\ref{KDENormalizationEq}.
\begin{eqnarray}
D_{KDE} = 1 - \frac{P_{KDE}}{P_{KDE} + 1/A} \label{KDENormalizationEq}
\end{eqnarray}

In Eq.~\ref{KDENormalizationEq}, $D_{KDE}$ is the normalized KDE distance and $P_{KDE}$ is the KDE
probability estimate.  $A$ is the area in state space covered by the training set: $A =
\prod\limits_{i} (q_{i, max} - q_{i, min})$.  This normalization therefore compares how likely a
point is given the KDE probability distribution estimate versus a uniform distribution.  As with the
normalized Mahalanobis distance, this normalized KDE distance varies from 0 (no extrapolation) to 1
(high extrapolation).

The trade-off between the Mahalanobis distance and the KDE distance is between memory efficiency and
flexibility.  While the Mahalanobis distance is more memory efficient--it only requires retaining
the mean and covariance matrix of the training data--it also makes strong assumptions about the
Gaussian nature of the training data.  The KDE distance requires much more memory usage, since it
stores the convolution of the Gaussian kernel with the entire training set.  However, it is able to
account for strongly non-Gaussian training data distributions. Therefore, a key question addressed by this paper is whether the Mahalanobis distance is an effective extrapolation metric, or whether the Gaussian assumption undermines its efficiency. The KDE method , which does not make this assumption, is less memory efficient and more computationally costly.

In this paper we will investigate the effectiveness of these two extrapolation metrics in detecting
regions of extrapolation for data-driven turbulence closures. It should be noted that the
calculations of these two extrapolation metrics only involve the RANS simulated mean flow
field. This is because the high fidelity data are usually unavailable for the flow to
  be predicted. For the RANS simulations, the same RANS model is used for both the training flows
  and the test flow, and the extrapolation metrics are calculated based on these RANS simulations.
Compared to DNS/LES, RANS simulations may fail to detect a flow feature and miss the
  correct flow topology. For instance, the separation would be falsely suppressed with a much
  shallower hill for the flow over periodic hills. Assuming shallower hill in both the training flow
  and the test flow, a similar suppression effect would exist for both flows if $k$-$\varepsilon$
  model is used, and the closeness between these two flows can still be detected. However, if such
  suppression only occurs in the training flows and is absent in the test flow, the extrapolation
  metrics would not suggest a close relationship between the training flows and the test flow, which
  can be seen as a conservative choice of estimating closeness between flow physics.
As we  used RANS simulations, the calculated extrapolation metrics would depend on the choice of RANS model. This impact of RANS model selection is especially desirable for the data-driven turbulence modeling, since the mean flow features used as inputs in the existing data-driven turbulence modeling~\cite{Wang2016,ling2016reynolds} are obtained based on RANS simulations. The extrapolation metrics are thus calculated based on the RANS simulations to assess the closeness of these inputs between the training flows and the test flow. In this work we explore the correlation of these extrapolation metrics with the machine learning prediction error for the test case where the high-fidelity data is available. By demonstrating such positive correlation, these two metrics can be used as indicators of the machine learning performance for the flows where the high-fidelity data are absent.

\section{Numerical Results}
\label{sec:results}

In this work, the extrapolation metrics based on Mahalanobis distance and kernel density estimation
are assessed for a given test set and several different training sets. The relationship between
machine learning prediction performance and these extrapolation metrics is also studied. The flow
over periodic hills~\cite{breuer2009flow} at $Re=10595$ is chosen as the test set. As shown in
Section~\ref{sec:extrapolation}, seven different training sets are employed in this work. The flow
configurations of training sets are illustrated in Fig.~\ref{fig:case-intro}. It should be noted
that only the lower part ($y/H<1.2$) of the training set is used for the random forest regression
due to the available high fidelity data from training sets. For all the baseline RANS simulations,
Launder-Sharma $k$-$\epsilon$ model~\cite{launder1974application} is used. It should
  be noted that the method proposed in this work is also directly applicable to other RANS
  models. However, the same RANS model needs to be employed for both the calculation of
  extrapolation metrics and the data-driven turbulence modeling procedure to ensure consistency.
The $y^+$ of the first cell center is kept less than 1 and thus no wall model is
applied.
The RANS simulations are performed in an open-source CFD platform OpenFOAM, using a
  built-in steady-state incompressible flow solver \texttt{simpleFoam}~\citep{weller1998tensorial},
  in which the SIMPLE algorithm is used. We choose the steady-state solver because the flow problems
  investigated in this work are all steady-state problems. For the numerical schemes, second-order
  central difference scheme is chosen for all terms except for the convection term, which is
  discretized with second-order upwind scheme.

%\subsection{Results}
%\textcolor{blue}{General point: 1. Demonstrate that the prediction performance is good if the training set is closely related to the prediction set. 2. Provides a quantitative approach to assess the closeness between different cases in feature space.}

\begin{figure}[!htbp]
  \centering
  \includegraphics[width=0.7\textwidth]{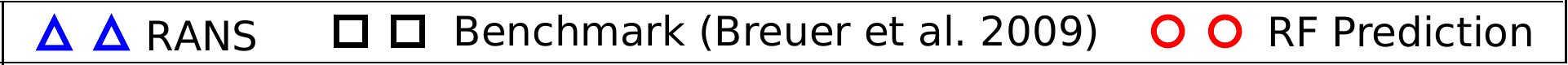}\\
  \subfloat[Training set: PH5600]{\includegraphics[width=0.4\textwidth]{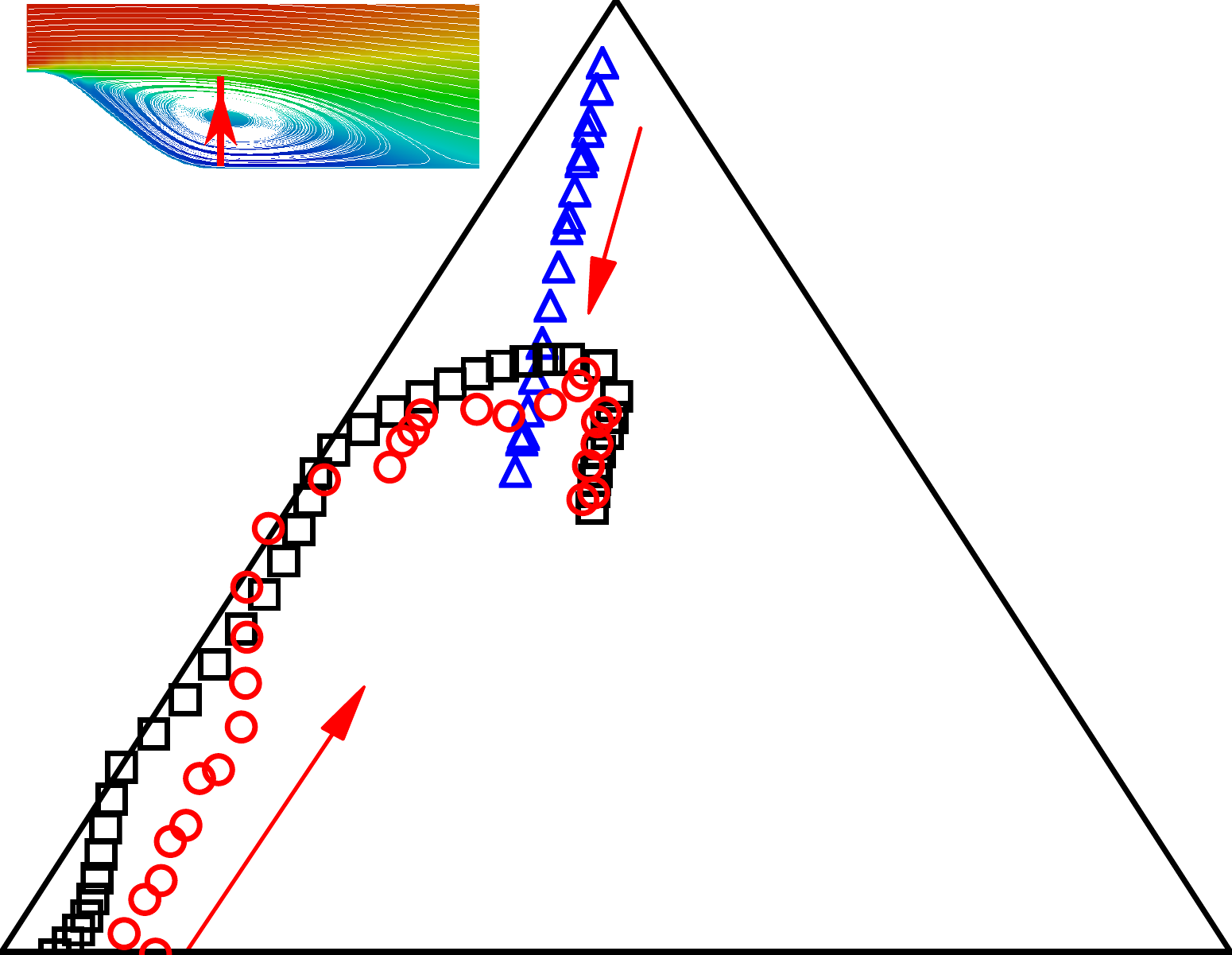}}\hspace{0.15em}
  \subfloat[Training set: PH1400]{\includegraphics[width=0.4\textwidth]{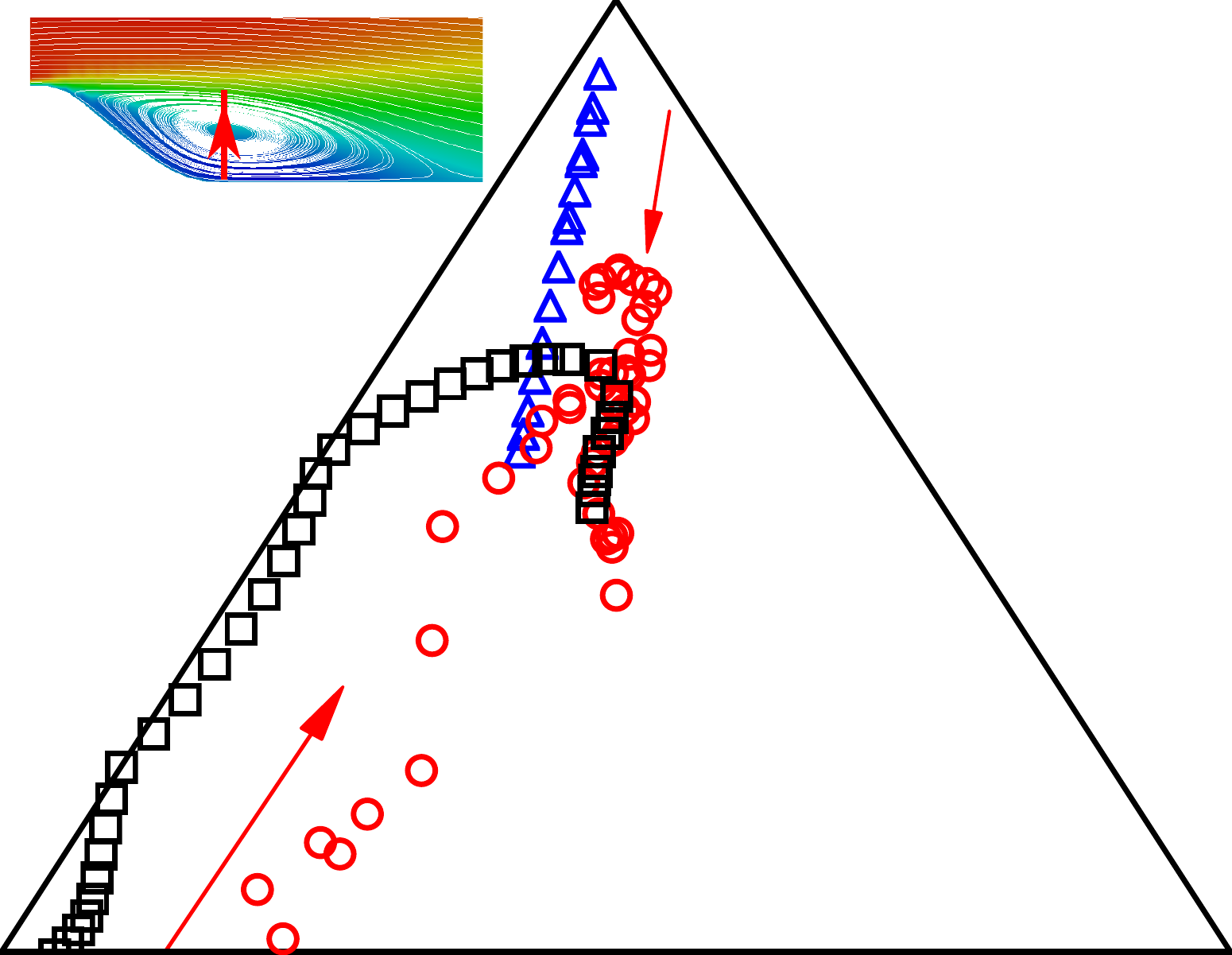}}\\
  \subfloat[Training set: CBFS13200]{\includegraphics[width=0.4\textwidth]{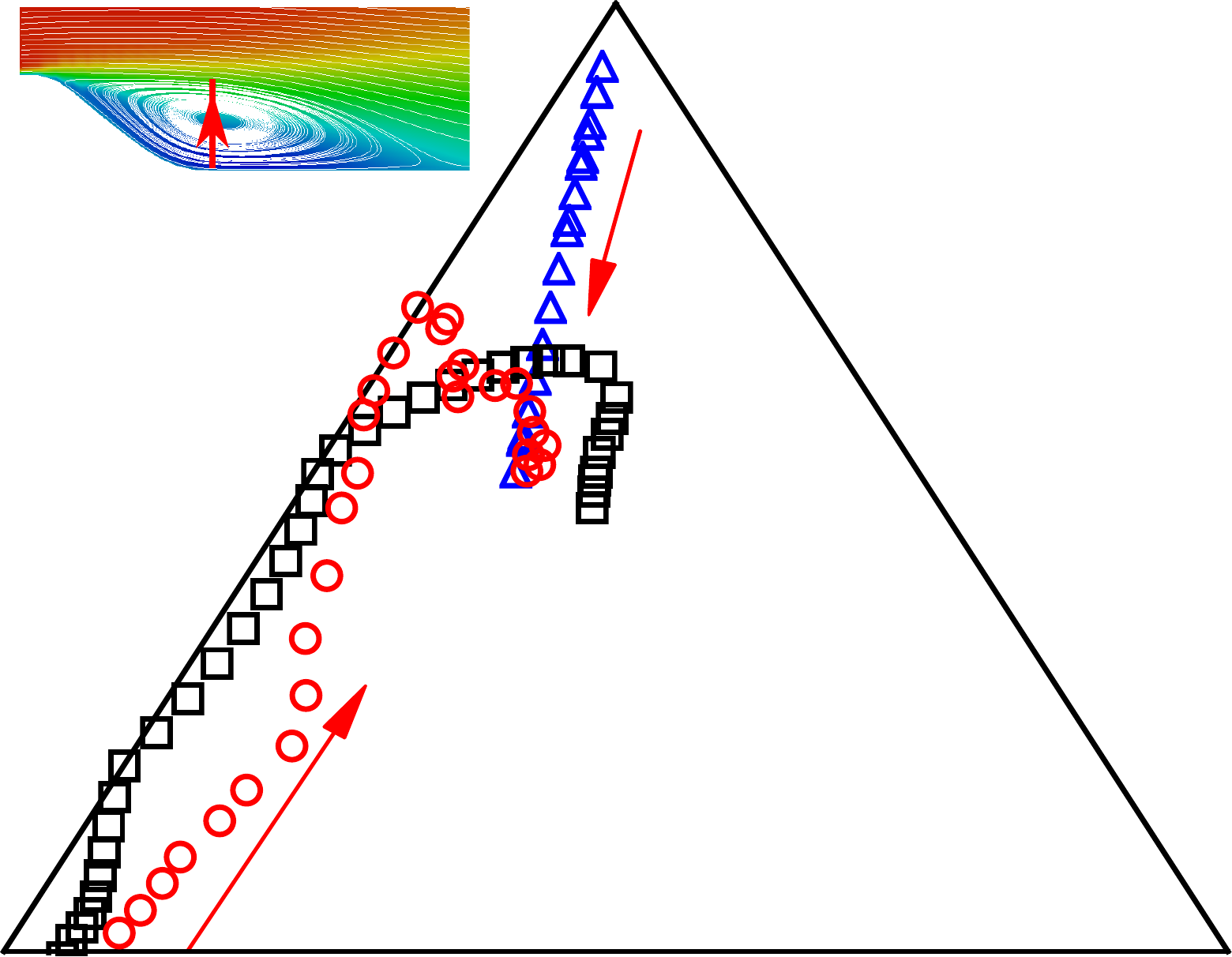}}\hspace{0.15em}
  \subfloat[Training set: WC360]{\includegraphics[width=0.4\textwidth]{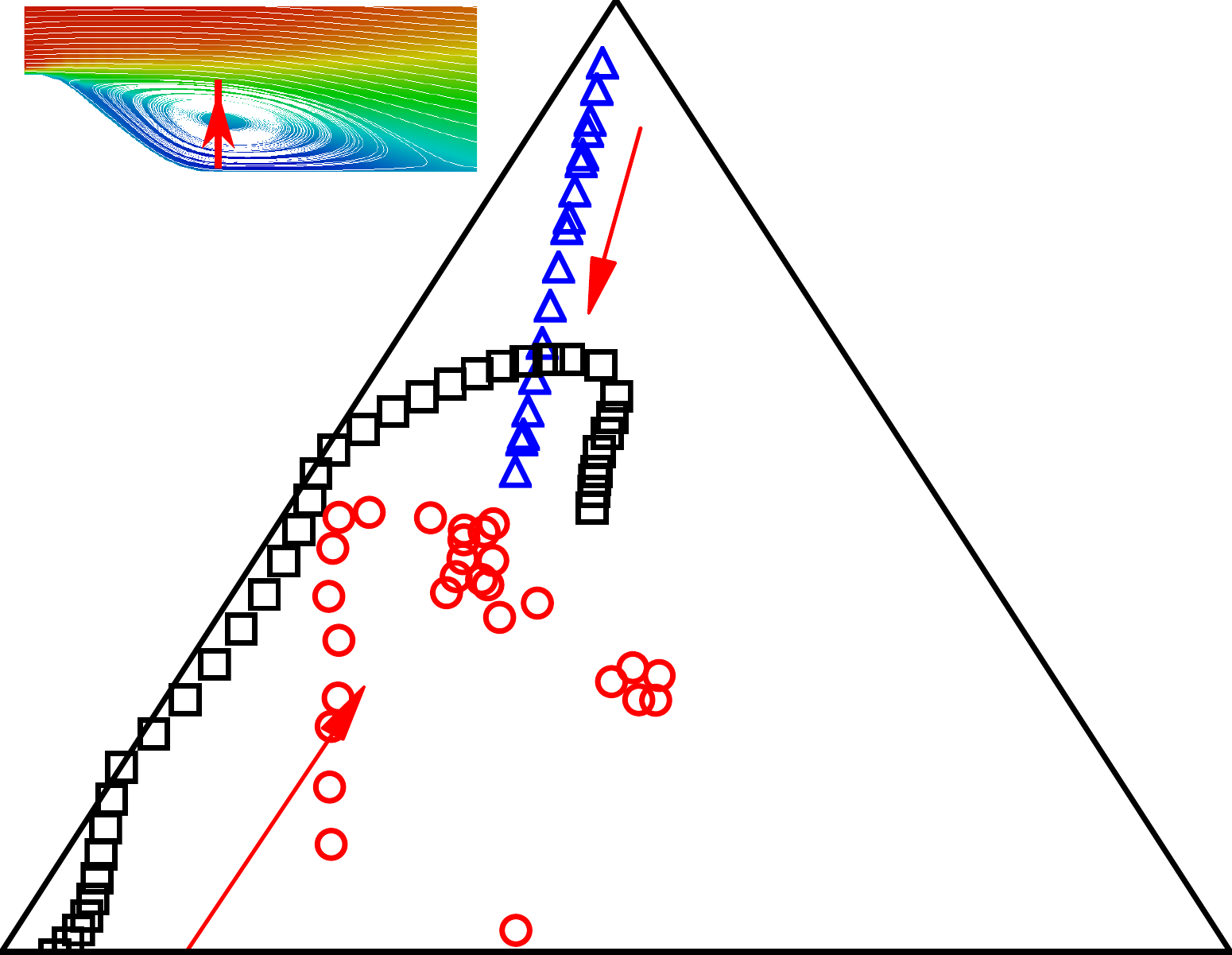}}\\
  \caption{Prediction of Reynolds stress anisotropy for the flow over periodic hills at $Re=10595$ along the line at $x/H=2$. The panels show the Reynolds stress anisotropy (a) with the flow over periodic hills at $Re=5600$ (PH5600) as training set, (b) with the flow over periodic hills at $Re=1400$ (PH1400) as training set, (c) with the flow over curved backward facing step (CBFS13200) as training set and (d) with the flow in wavy channel (WC360) as training set.}
\label{fig:bary2}
\end{figure}

\begin{figure}[!htbp]
  \centering
  \includegraphics[width=0.7\textwidth]{baryPlot-legend}\\
  \subfloat[Training set: PH5600]{\includegraphics[width=0.4\textwidth]{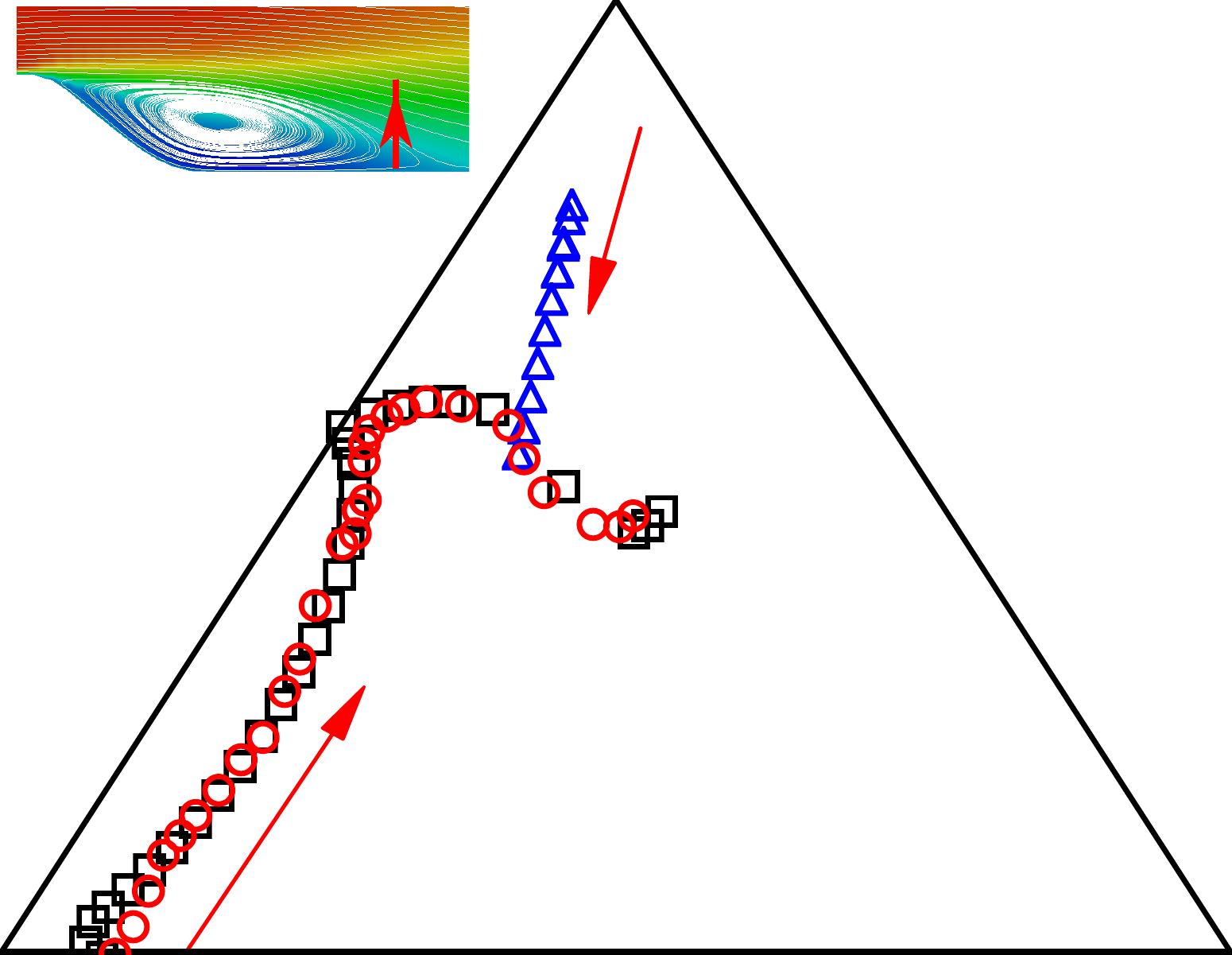}}\hspace{0.15em}
  \subfloat[Training set: PH1400]{\includegraphics[width=0.4\textwidth]{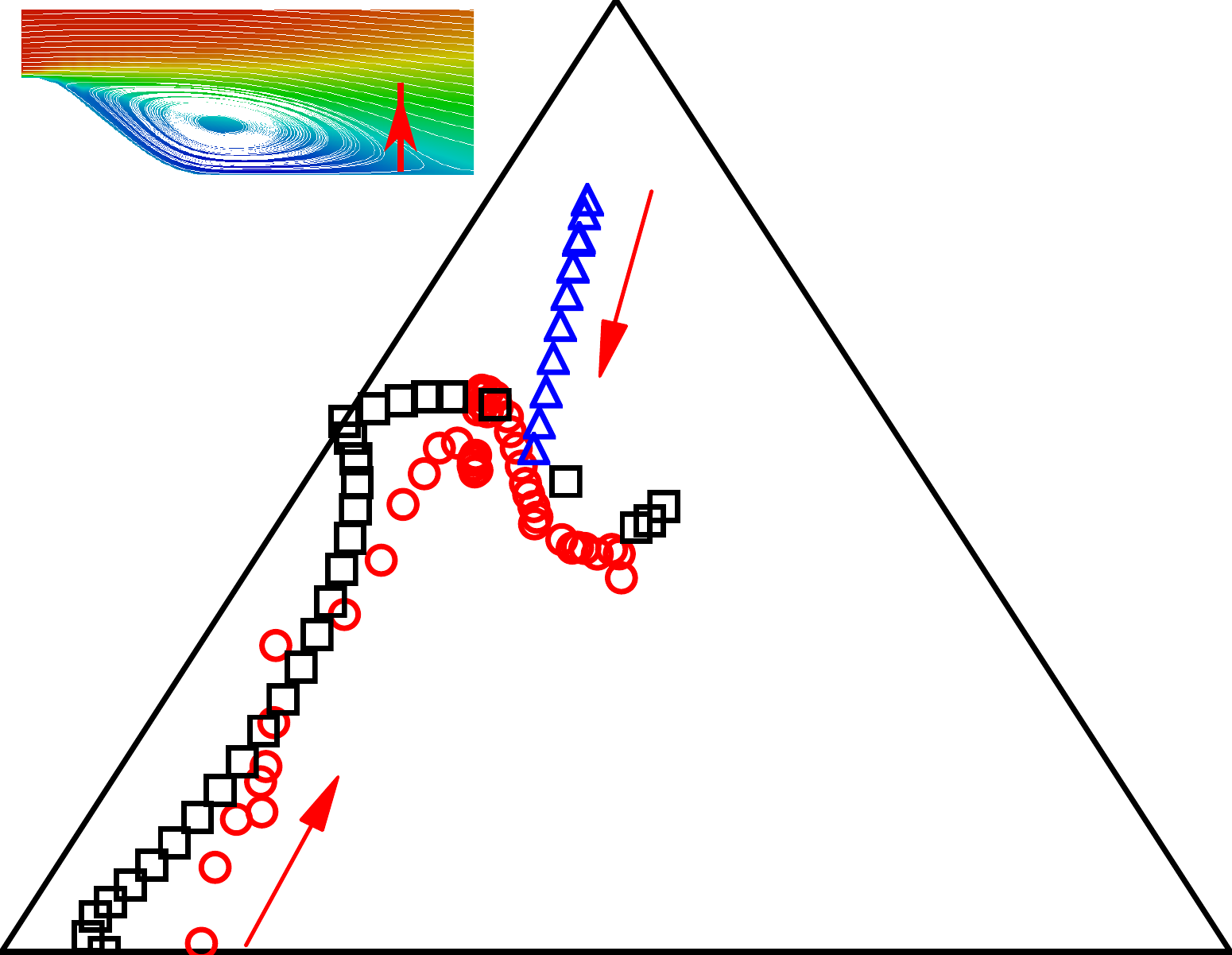}}\\
  \subfloat[Training set: CBFS13200]{\includegraphics[width=0.4\textwidth]{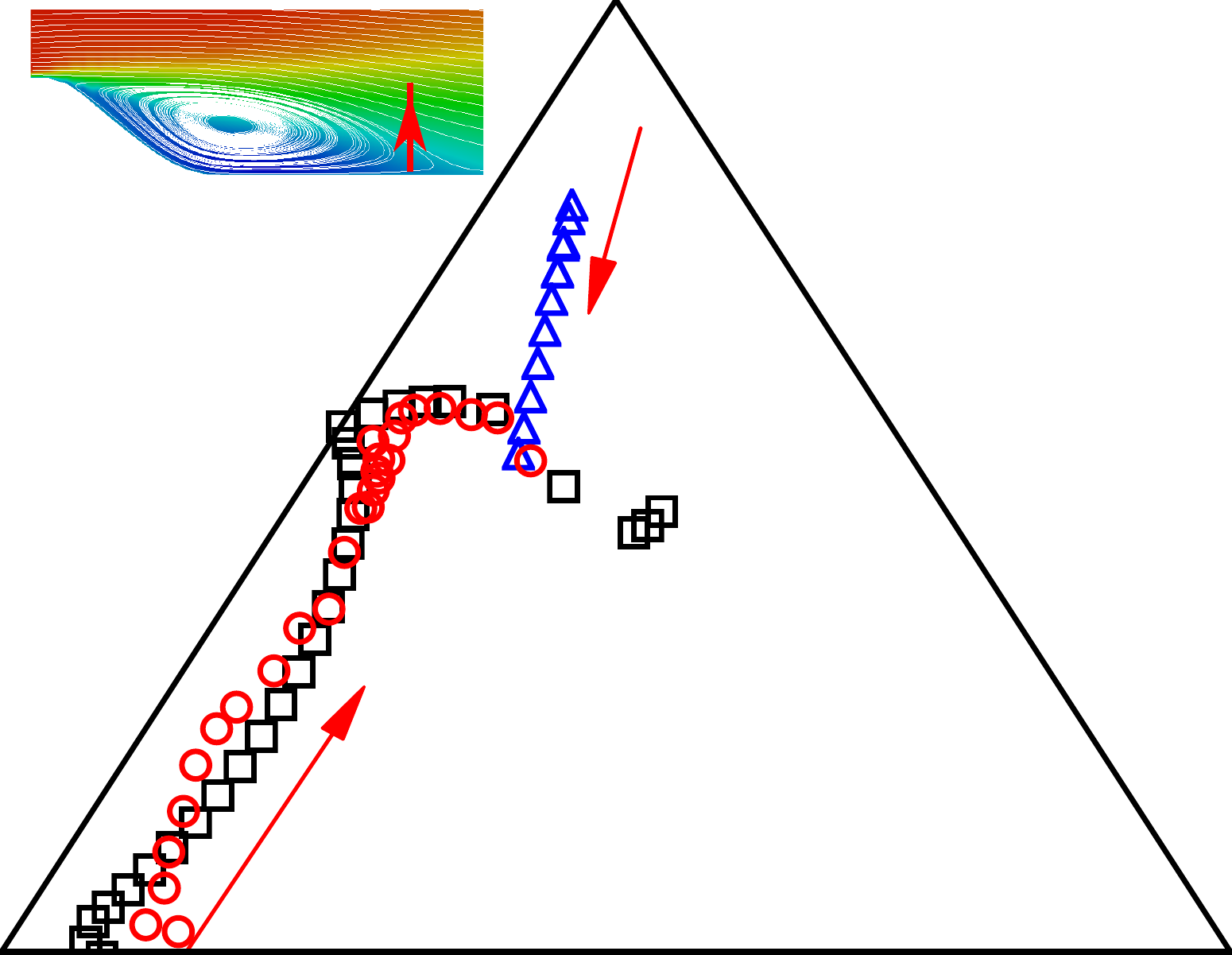}}\hspace{0.15em}
  \subfloat[Training set: WC360]{\includegraphics[width=0.4\textwidth]{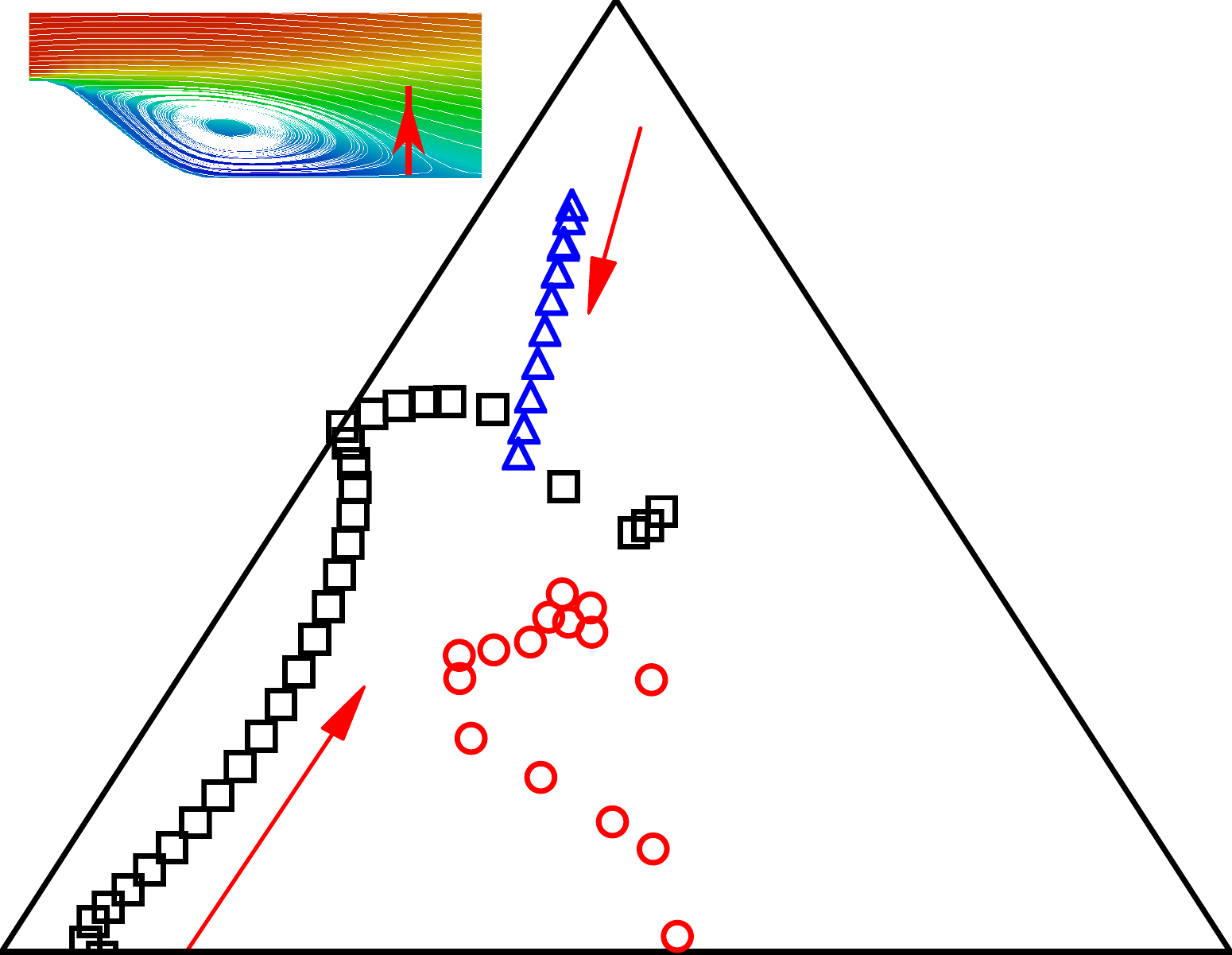}}\\
  \caption{Prediction of Reynolds stress anisotropy for the flow over periodic hills at $Re=10595$ along the line at $x/H=4$. The panels show the Reynolds stress anisotropy (a) with the flow over periodic hills at $Re=5600$ (PH5600) as training set, (b) with the flow over periodic hills at $Re=1400$ (PH1400) as training set, (c) with the flow over curved backward facing step (CBFS13200) as training set and (d) with the flow in wavy channel (WC360) as training set.}
\label{fig:bary4}
\end{figure}

The random forest regression is performed based on each of the training sets. The number of max features is set as 6, considering that there are 10 input features in this work. The number of trees is set as 100. This number is chosen by observing the out-of-bag (OOB) error to avoid possible overfitting on the training sets. We have observed the OOB error with different numbers of trees (50, 100, 150) and this error is not sensitive based on our current setting of the number of trees. Representative results for the Reynolds stress anisotropy from four of the training sets are shown in Figs.~\ref{fig:bary2} and~\ref{fig:bary4}. The Reynolds stress
anisotropy profiles were taken along two lines at $x/H=2$ and $x/H=4$, both of which are within the
recirculation region. It can be seen in Figs.~\ref{fig:bary2}(a) and~\ref{fig:bary4}(a) that the predicted Reynolds
stress anisotropies are in good agreement with the benchmark data if the flow over periodic hills at
$Re=5600$ is used as the training set. The prediction is less accurate but still captures the general pattern of benchmark data if the flow at $Re=1400$ is used as training set. If the training set is the flow over curved step, the
predicted Reynolds stress anisotropy is still significantly improved versus the default RANS predictions and the pattern of the
benchmark data is generally predicted as shown in Figs.~\ref{fig:bary2}(c) and~\ref{fig:bary4}(c). The predicted Reynolds stress
anisotropy state is significantly less accurate if the training set is the flow in a channel with a wavy bottom wall as shown in Figs.~\ref{fig:bary2}(d) and~\ref{fig:bary4}(d). In this work, these quantities that directly predicted by the trained discrepancy functions are compared with the extrapolation metrics. The reconstructed Reynolds stress components can be found in a separate work by Wang et al.~\cite{Wang2016}. It should be noted that the specification of DNS Reynolds stress into RANS equations would reduce the robustness of the numerical simulation~\cite{thompson2016methodology,poroseva2016accuracy}. For the data-driven turbulence modeling, some attempts to propagate the predicted Reynolds stress to the mean velocity via RANS equations can be found in the work by Wang et al.~\cite{Wang2017}. The comparison in
Figs.~\ref{fig:bary2} and~\ref{fig:bary4} indicates that the prediction performance
depends on the choice of the training set. If the training set is closely related to the
test set, e.g., the geometry is the same and the Reynolds number is slightly different, the
predicted Reynolds stress anisotropy is more reliable. To assess the confidence of the
prediction in the practical scenario where benchmark data are not available, the closeness
between different flows needs to be defined.

\begin{figure}[!htbp]
  \centering
  \subfloat[Reynolds number extrapolation]{\includegraphics[width=0.495\textwidth]{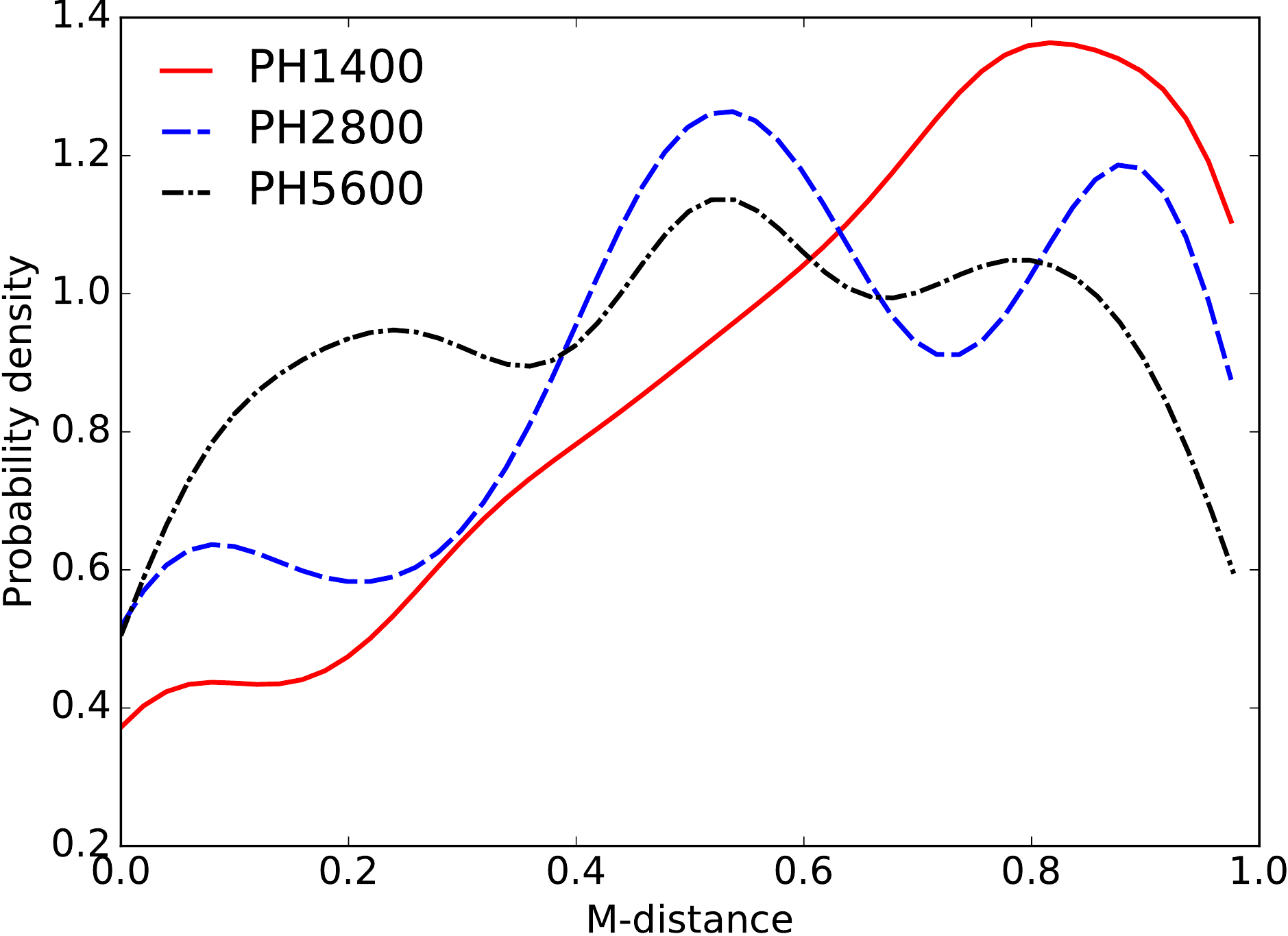}}\hspace{0.1em}
  \subfloat[Geometry extrapolation]{\includegraphics[width=0.495\textwidth]{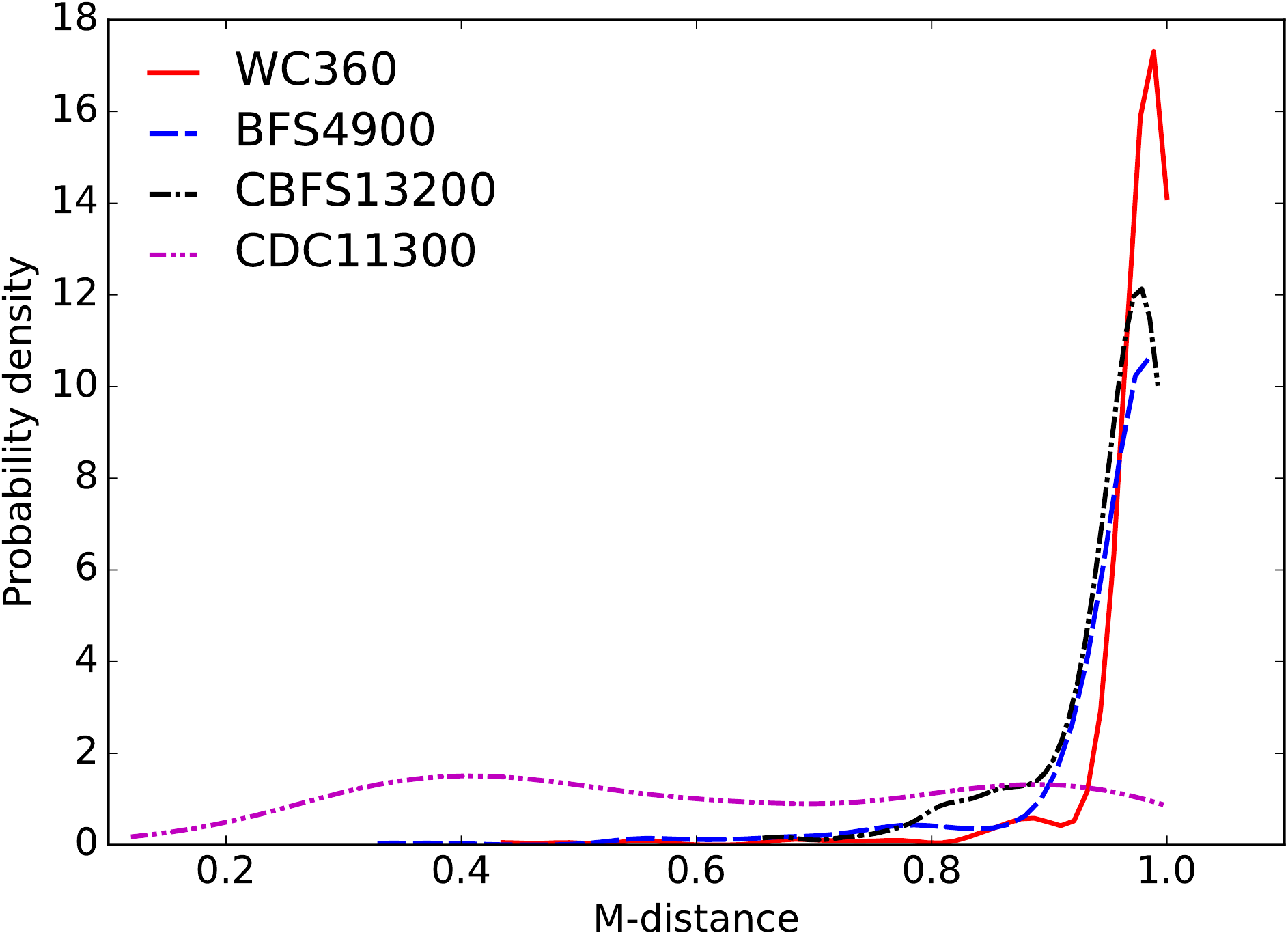}}\\
  \caption{Probability density function of Mahalanobis distance based on different training
    sets. (a) The distribution of Mahalanobis distance based on Reynolds number extrapolation. (b)
    The distribution of Mahalanobis distance based on geometry extrapolation. All the Mahalanobis
    distances have been normalized into the range from zero to one, as detailed in Section~\ref{sec:extrapolation}.}
\label{fig:Mdist-pdf}
\end{figure}

We use the Mahalanobis distance and the KDE distance as extrapolation metrics to
gauge the closeness between different flows in feature space. The distribution of normalized Mahalanobis
distance between the flow over periodic hills at $Re=10595$ and each training set is shown in
Fig.~\ref{fig:Mdist-pdf}. For the training set with same geometry but
different Reynolds numbers, it can be seen in Fig.~\ref{fig:Mdist-pdf} that the mean Mahalanobis
distance increases as the difference between the Reynolds numbers becomes larger. This is consistent with
the empirical knowledge that flows with same geometric configuration are often closely related if the Reynolds
number difference is small.

% \todo[inline]{JL: add the anisotropy prediction from Re=2800/1400}

For the training sets with a different geometry, it can be seen in Fig.~\ref{fig:Mdist-pdf}(b) that
the Mahalanobis distances are generally greater than that shown in Fig.~\ref{fig:Mdist-pdf}(a). In
addition, the Mahalanobis distances based on the flow in a wavy channel are generally larger than
the Mahalanobis distances based on the flow over curved step. Compared with the prediction
performance as shown in Figs.~\ref{fig:bary2} and~\ref{fig:bary4}, this suggests that the Mahalanobis distance can be an
indicator to estimate the prediction performance when the benchmark data are not available. In
practice, the normalized Mahalanobis distances can be obtained \textit{a priori} based on the mean
flow features from training set and test set. Generally speaking, smaller mean Mahalanobis 
distance indicates that the RANS simulation of the training set is more similar to the RANS 
simulation of the test set. Assuming that similar RANS simulation results indicate similar flow physics, better prediction performance can be expected.

\begin{figure}[!htbp]
  \centering
  \subfloat[Reynolds number extrapolation]{\includegraphics[width=0.495\textwidth]{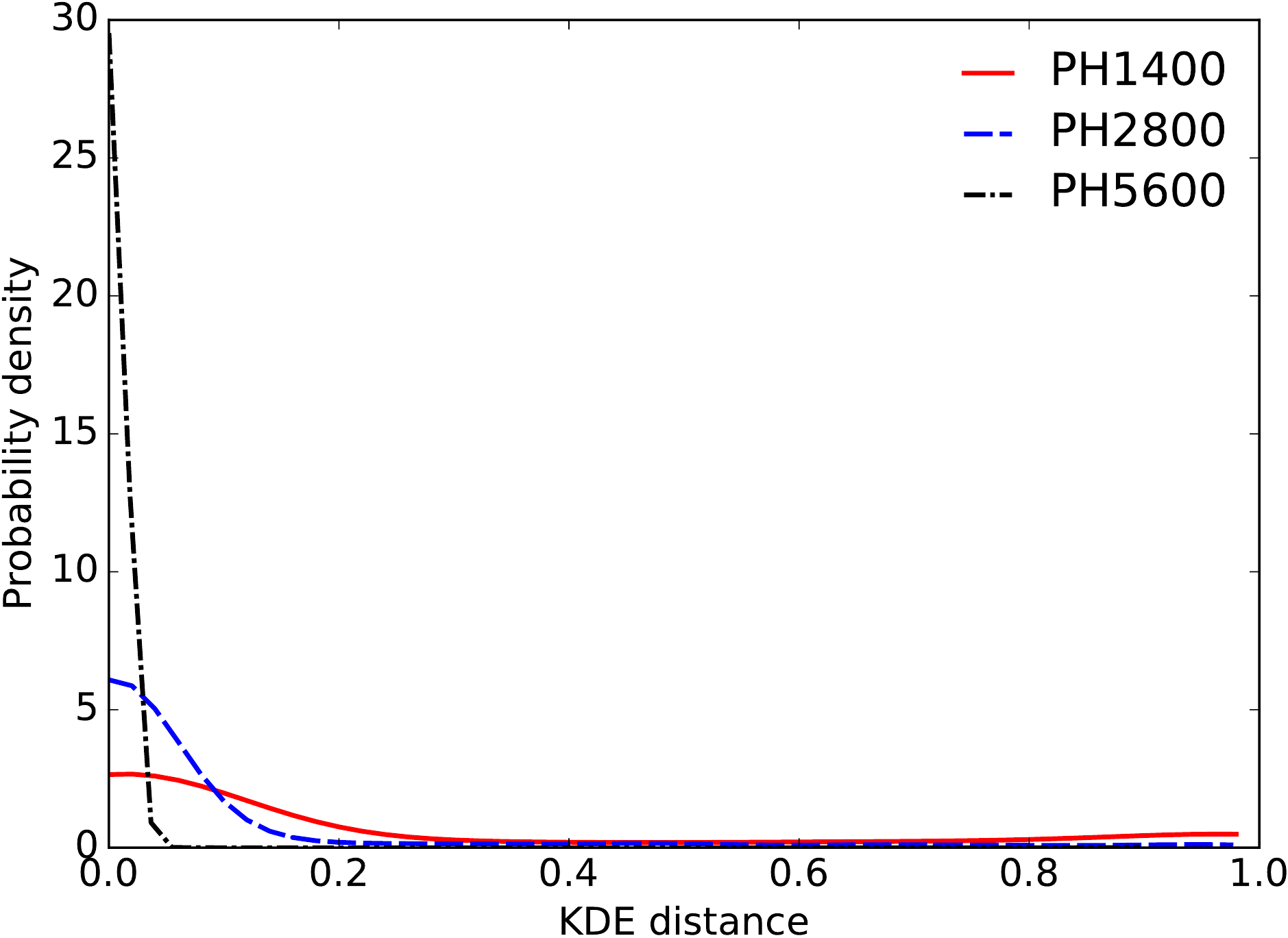}}\hspace{0.1em}
  \subfloat[Geometry extrapolation]{\includegraphics[width=0.495\textwidth]{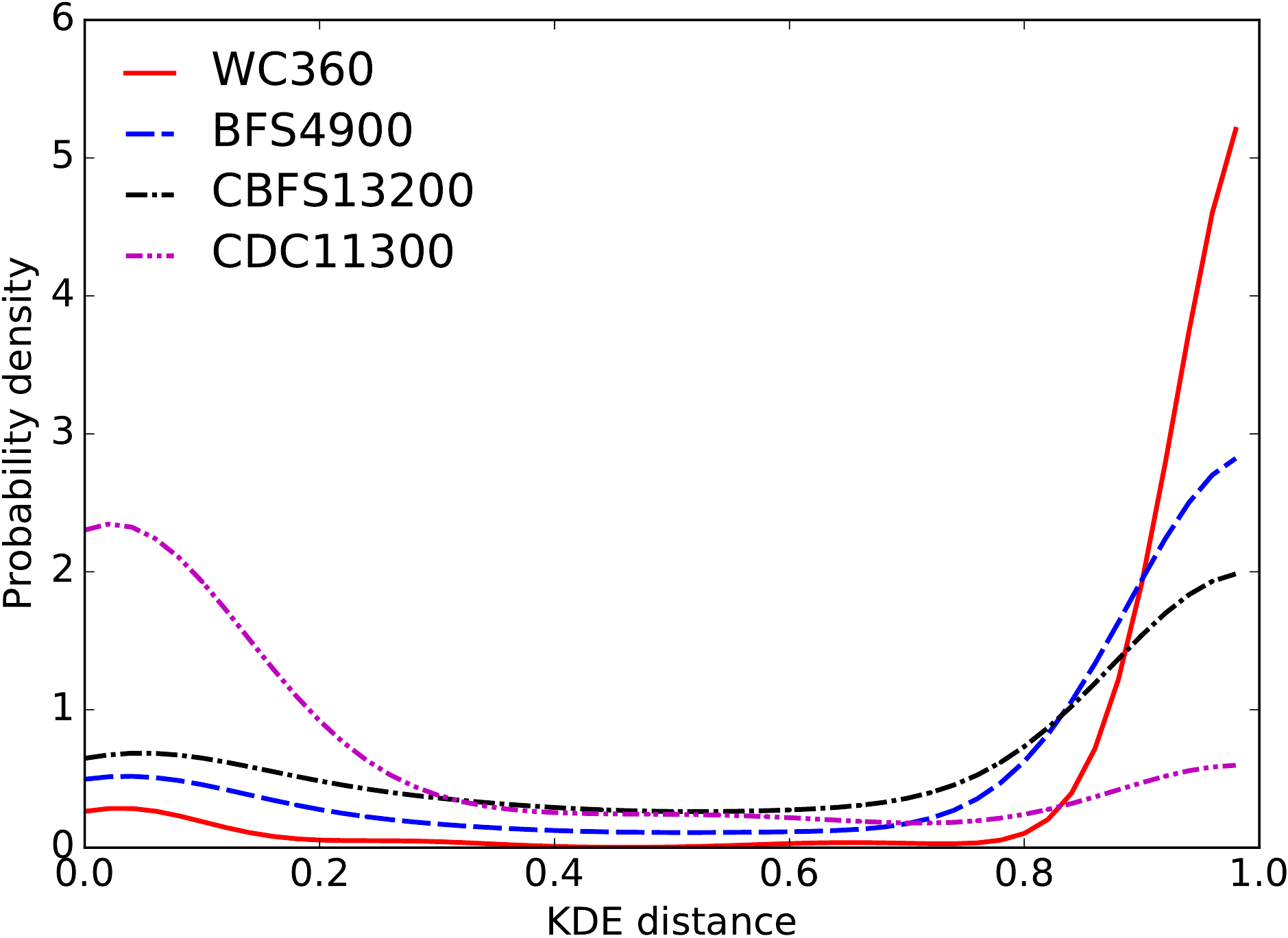}}\\
  \caption{Probability density function of KDE distance based on different training
    sets. (a) The distribution of KDE distance based on Reynolds number extrapolation. (b)
    The distribution of KDE distance based on geometry extrapolation. All the KDE
    distances have been normalized by the estimated area occupied by training set in feature space.}
\label{fig:kde-dist-pdf}
\end{figure}

Similar to the Mahalanobis distance, the KDE distances are generally smaller for Reynolds number extrapolation as shown in Fig.~\ref{fig:kde-dist-pdf}(a), compared with the KDE distances based on geometry extrapolation as shown in Fig.~\ref{fig:kde-dist-pdf}(b). This is consistent with the Mahalanobis distance results as shown in Fig.~\ref{fig:Mdist-pdf}, indicating that Mahalanobis distance can provide an overall reliable extrapolation estimation in spite of its Gaussian distribution assumption. 

% \textcolor{blue}{Main point 3: The prediction performance is better for the extrapolation between
% different Reynolds numbers. It indicates that better prediction performance can be achieved if the
% mean flow feature between the training set and prediction set is close. However, such observation
% does not hold true for the extrapolation between different geometries. Although the separation
% bubble of converging-diverging channel is significantly different from the periodic hill case, the
% prediction is better than the wavy channel case and the backward step case. }

\begin{figure}[!htbp]
  \centering
  \subfloat[Mahalanobis distance]{\includegraphics[width=0.495\textwidth]{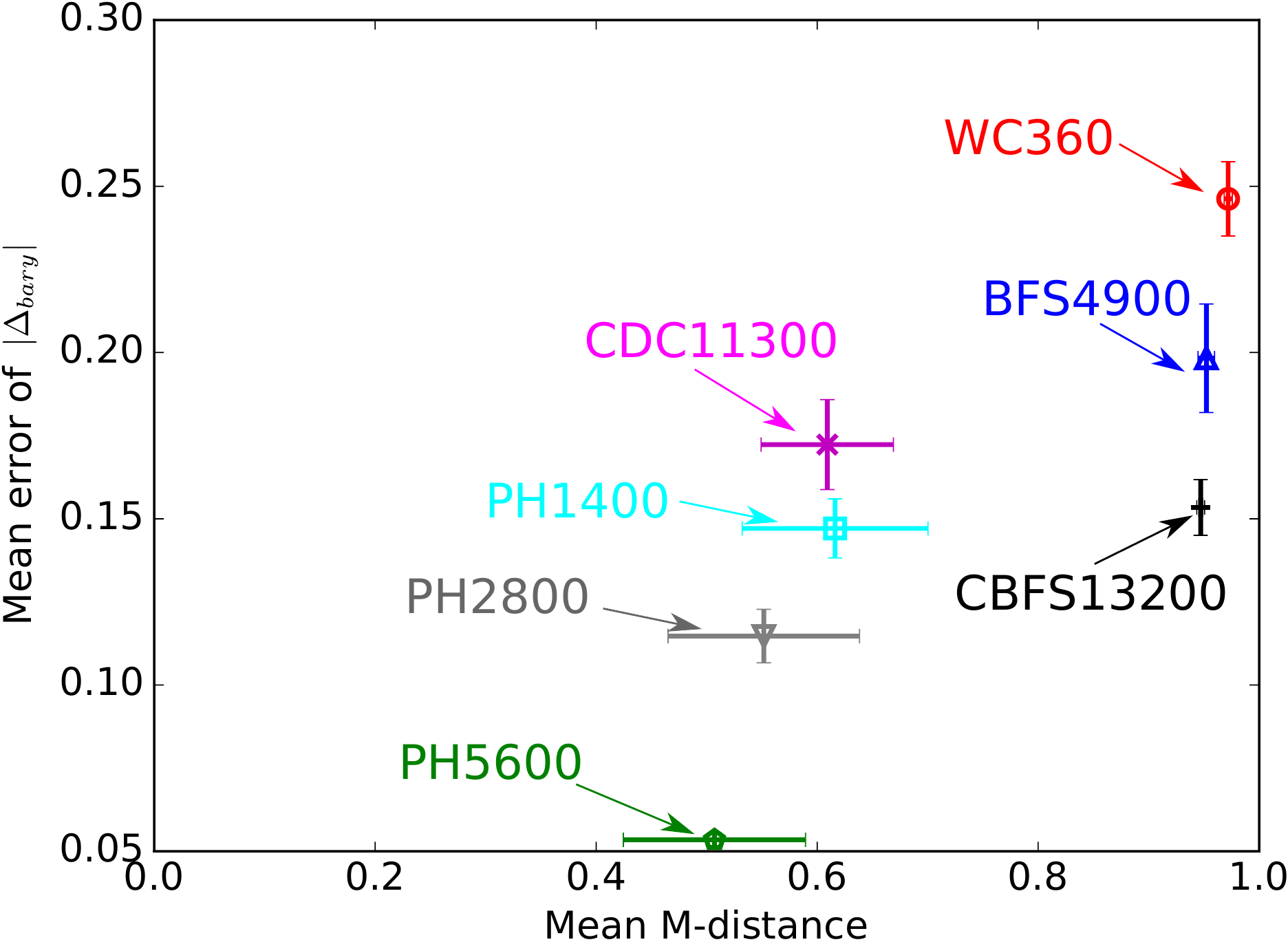}}\hspace{0.1em}
  \subfloat[KDE distance]{\includegraphics[width=0.495\textwidth]{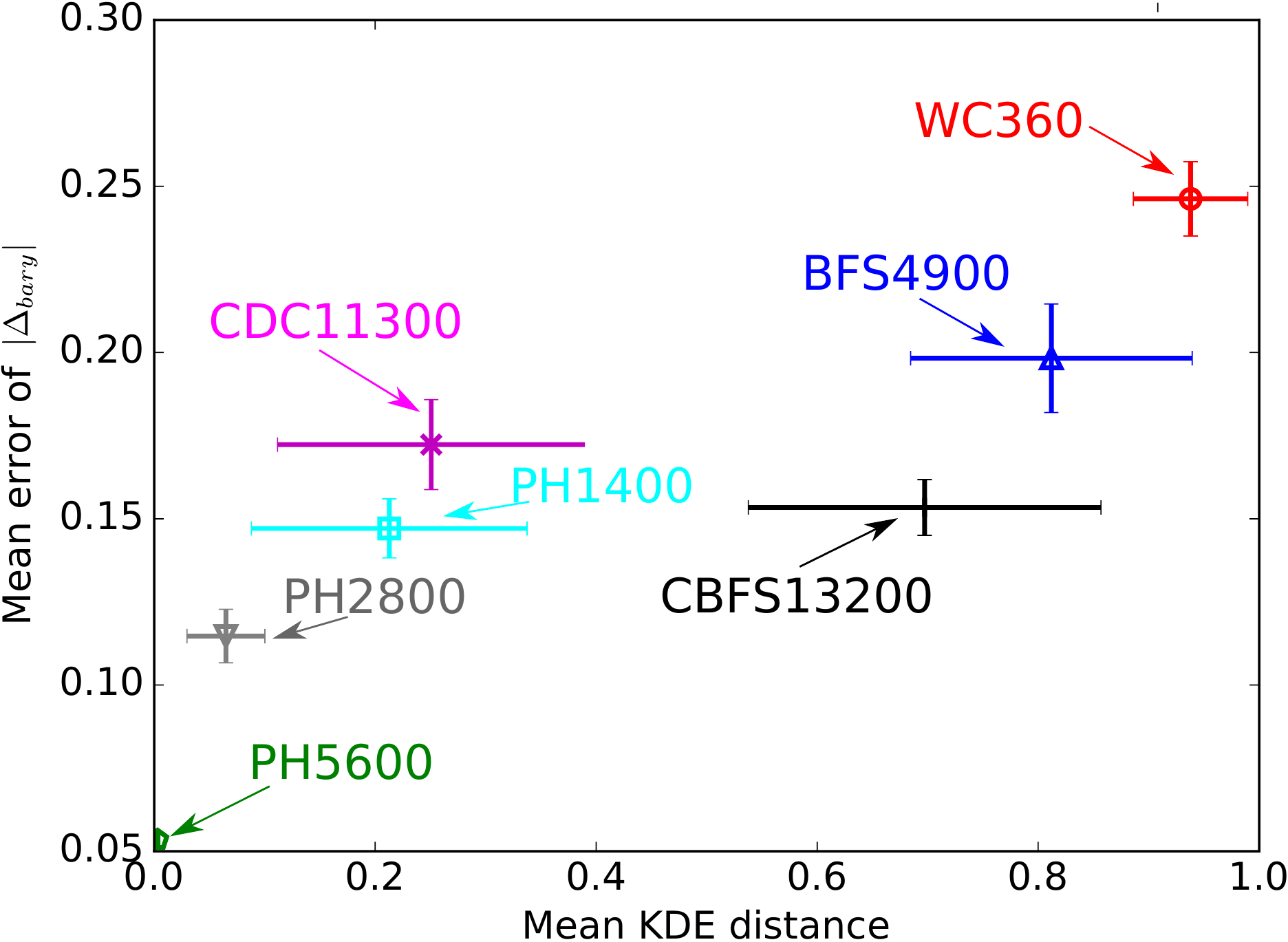}}
  \caption{ The relationship between the mean error of the
    prediction of Reynolds stress anisotropy and the mean value of extrapolation metrics, i.e., (a) Mahalanobis distance and (b) KDE distance. The standard deviations of both extrapolation metrics are shown as the horizontal bars. The standard deviations of the prediction error of anisotropy are also shown as the vertical bars.}
\label{fig:MSE}
\end{figure}

By analyzing the mean prediction error of Reynolds stress anisotropy based on different training sets, we can see that there is a positive correlation
between mean prediction error of Reynolds stress anisotropy and the extrapolation metrics as shown in Fig.~\ref{fig:MSE}. Figure~\ref{fig:MSE} also shows that the Mahalanobis distance based on geometry extrapolation is close to one and the standard deviation is much smaller than that based on Reynolds number extrapolation. A possible explanation is that the normalization procedure may lead to a dense clustering near the value of one. Compared to the KDE distance as shown in Fig.~\ref{fig:MSE}(b), it can be seen in Fig.~\ref{fig:MSE}(a) that the trend of Mahalanobis distance is similar to that of KDE distance for different training sets. This result suggests that the Mahalanobis distance, while simpler, may still be effective as an extrapolation metric on turbulence data sets.

%\textcolor{blue}{Main point 4: The analysis of M-distance contour and the prediction performance for different training cases.}

\begin{figure}[!htbp]
  \centering
  \includegraphics[width=0.5\textwidth]{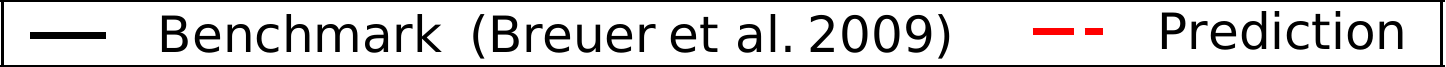}\\
  \subfloat[Prediction of $\xi$]{\includegraphics[width=0.495\textwidth]{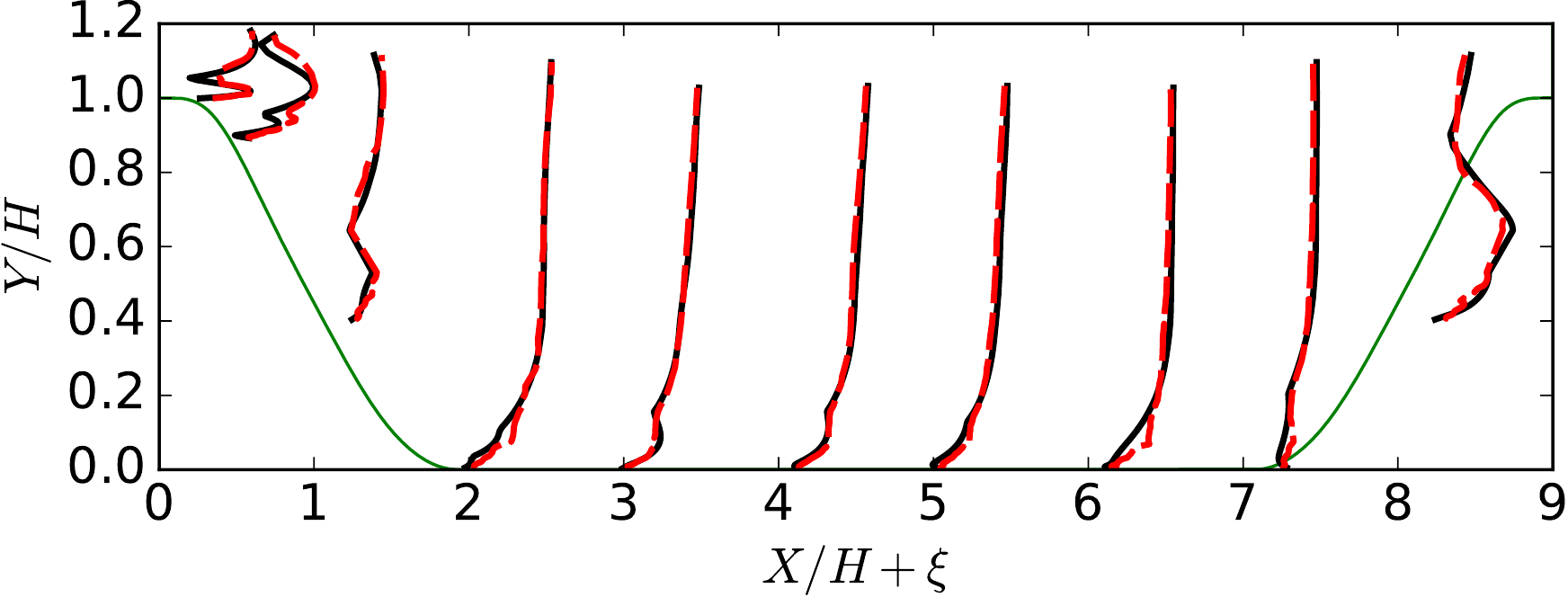}}\hspace{0.1em}
  \subfloat[Prediction of $\eta$]{\includegraphics[width=0.495\textwidth]{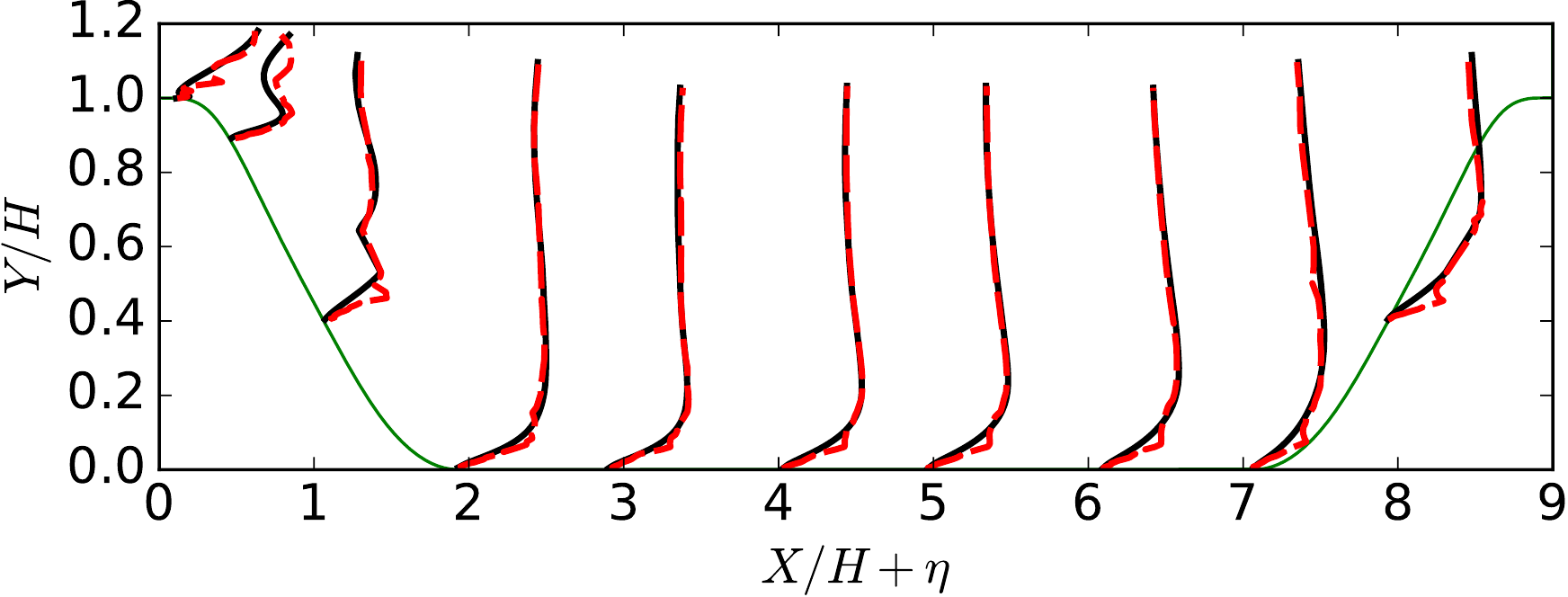}}\\
  \includegraphics[width=0.35\textwidth]{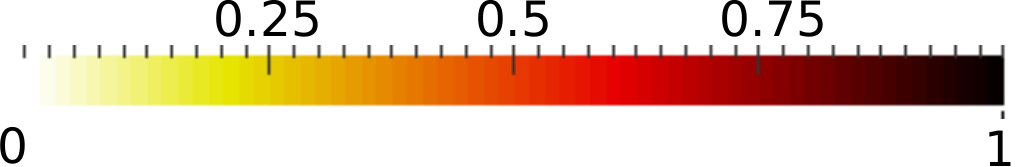}\\
  \vspace{-0.8em}
  \subfloat[Mahalanobis distance]{\includegraphics[width=0.45\textwidth, height=0.12\textwidth]{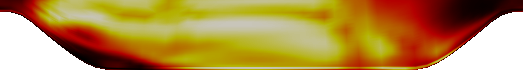}}\hspace{0.1em}
  \subfloat[KDE distance]{\includegraphics[width=0.45\textwidth, height=0.12\textwidth]{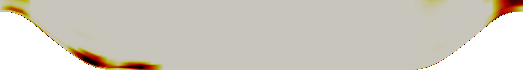}}\\
  \caption{Prediction of Reynolds stress anisotropy for the flow over periodic hills at $Re=10595$ along (a)
    horizontal direction $\xi$ and (b) vertical direction $\eta$ of Barycentric triangle. \textbf{The training set is the flow over
    periodic hills at \boldmath{$Re=5600$}}. The profiles are shown at 10 locations, $x/H$ = 0,
    0.5, 1, 2, ..., 8. The benchmark discrepancy is obtained based on the Reynolds stress from LES
    simulation~\cite{breuer2009flow}. The contours of (c) Mahalanobis distance and (d) KDE distance are presented for comparison.}
\label{fig:bary_Re5600}
\end{figure}

\begin{figure}[!htbp]
  \centering
  \includegraphics[width=0.5\textwidth]{pred-legend}\\
  \subfloat[Prediction of $\xi$]{\includegraphics[width=0.495\textwidth]{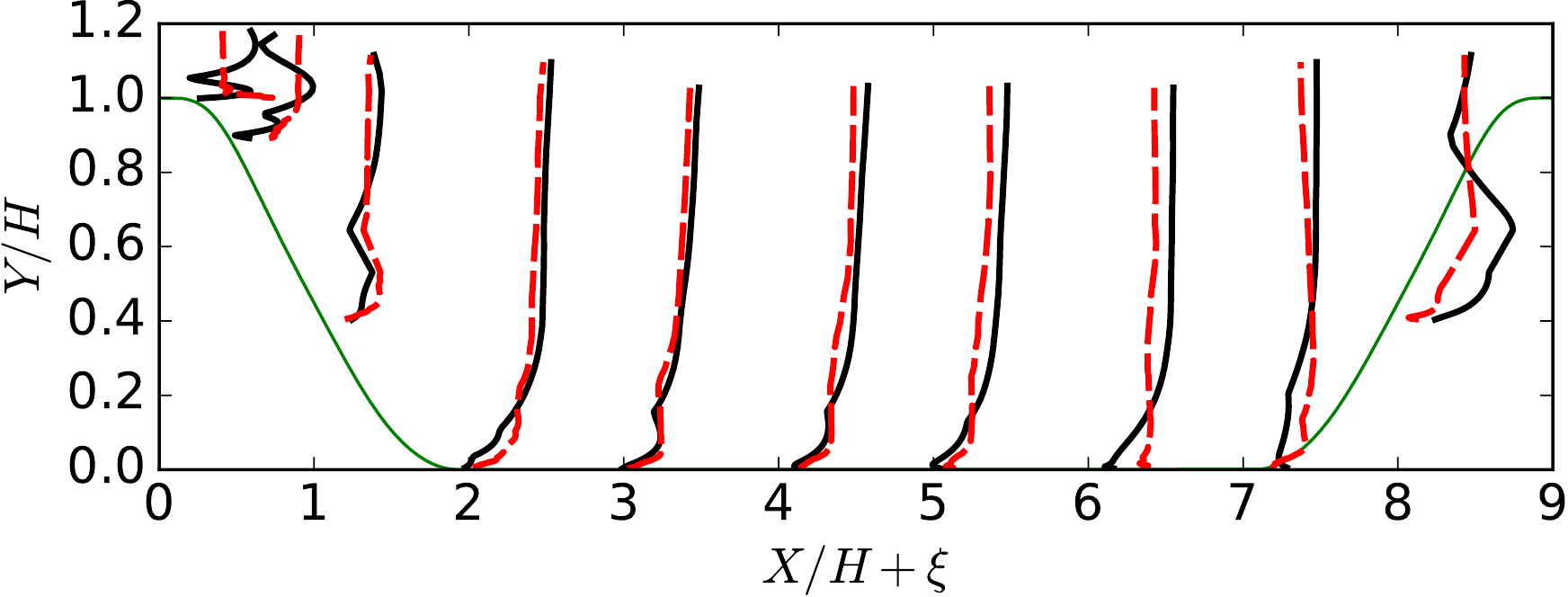}}\hspace{0.1em}
  \subfloat[Prediction of $\eta$]{\includegraphics[width=0.495\textwidth]{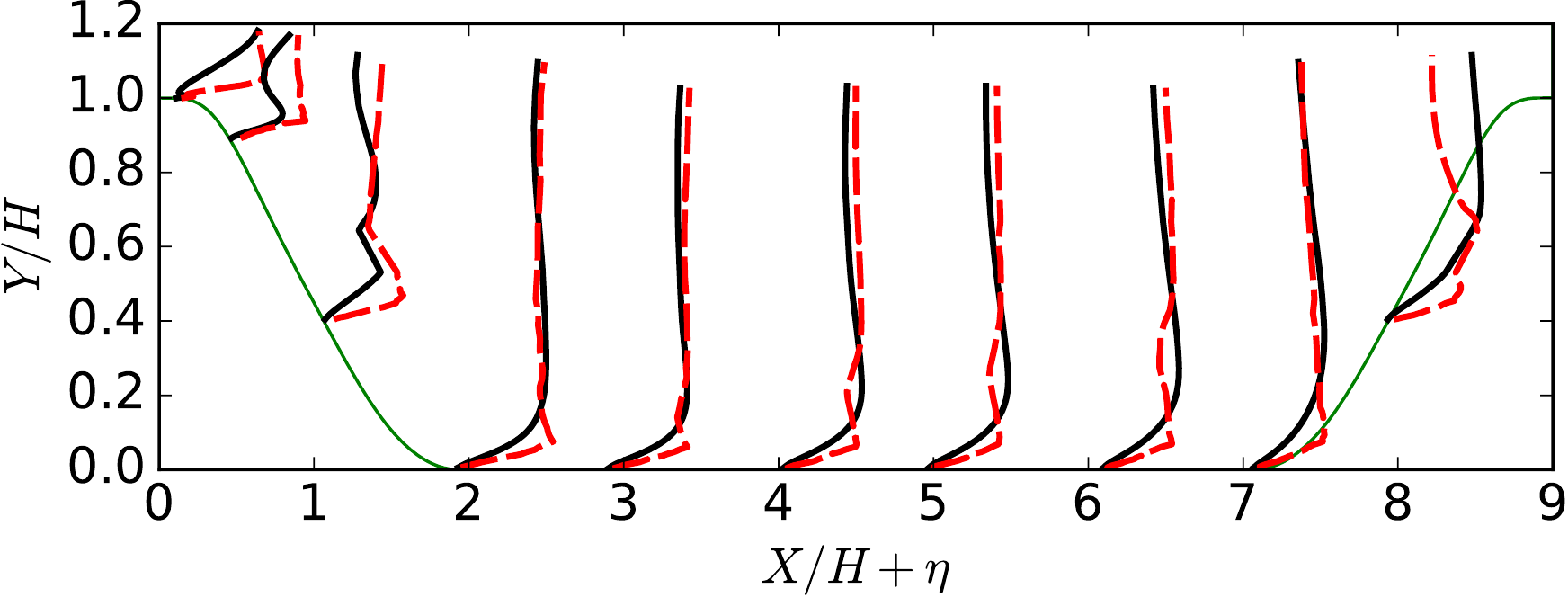}}\\
  \includegraphics[width=0.35\textwidth]{contour-legend}\\
  \vspace{-0.8em}
  \subfloat[Mahalanobis distance]{\includegraphics[width=0.45\textwidth, height=0.12\textwidth]{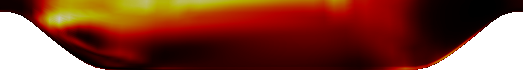}}\hspace{0.1em}
  \subfloat[KDE distance]{\includegraphics[width=0.45\textwidth, height=0.12\textwidth]{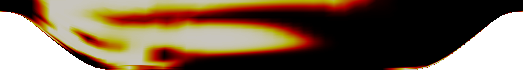}}\\
  \caption{Prediction of Reynolds stress anisotropy for the flow over periodic hills at $Re=10595$ along (a)
    horizontal direction $\xi$ and (b) vertical direction $\eta$ of Barycentric triangle. \textbf{The training set is the flow over
    curved backward facing step at \boldmath{$Re=13200$}}. The profiles are shown at 10 locations, $x/H$ = 0,
    0.5, 1, 2, ..., 8. The benchmark discrepancy is obtained based on the Reynolds stress from LES
    simulation~\cite{breuer2009flow}. The contours of (c) Mahalanobis distance and (d) KDE distance are presented for comparison.}
\label{fig:bary_CBFS}
\end{figure}

\begin{figure}[!htbp]
  \centering
  \includegraphics[width=0.5\textwidth]{pred-legend}\\
  \subfloat[Prediction of $\xi$]{\includegraphics[width=0.495\textwidth]{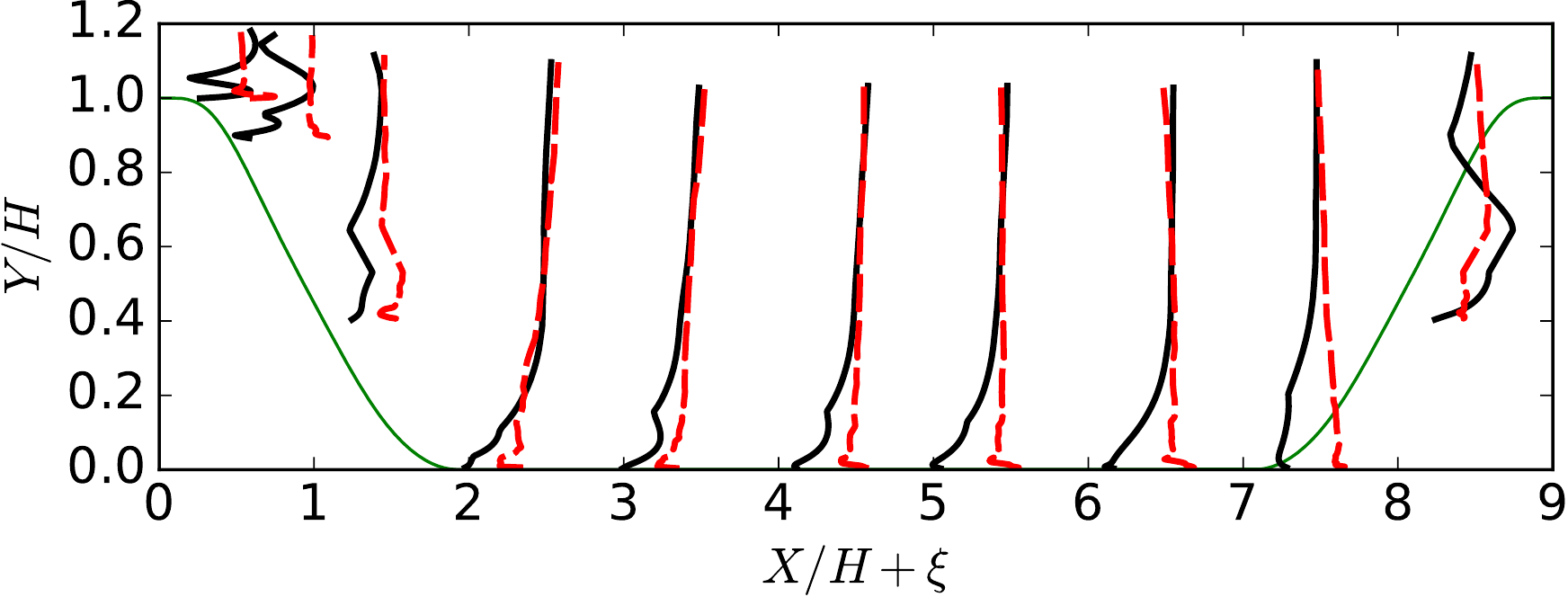}}\hspace{0.1em}
  \subfloat[Prediction of $\eta$]{\includegraphics[width=0.495\textwidth]{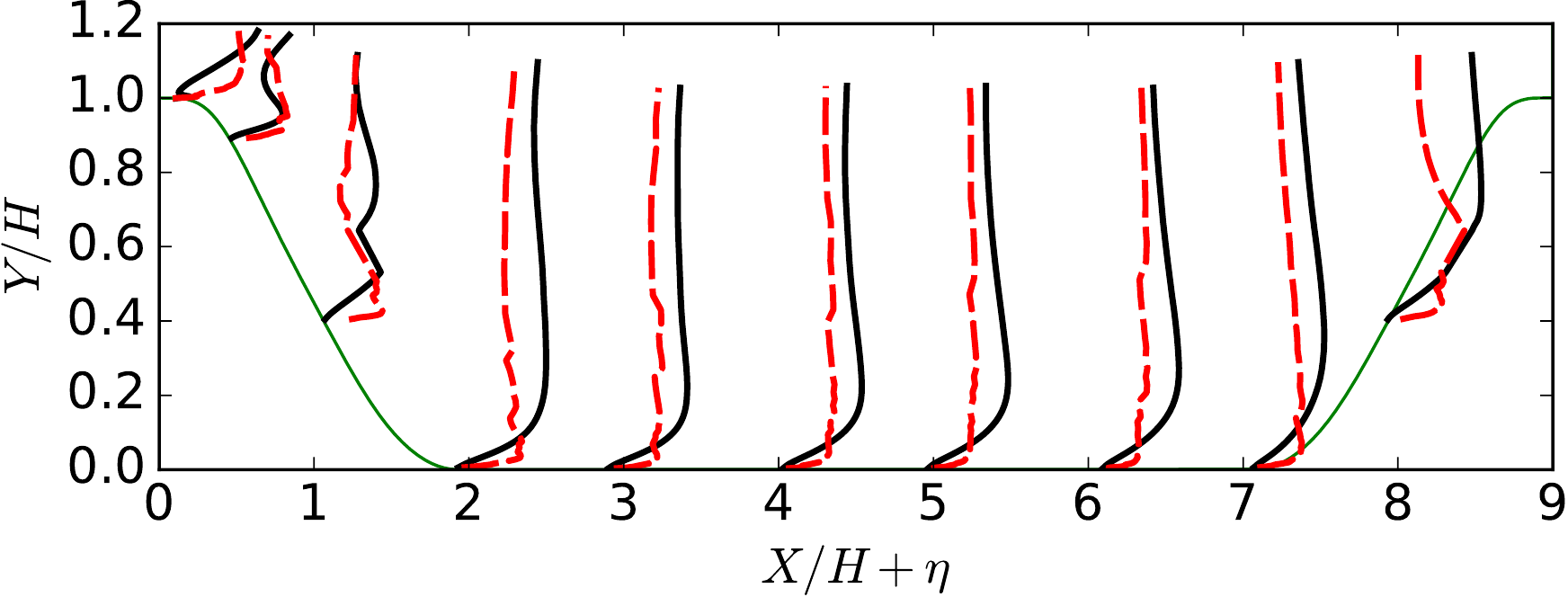}}\\
  \includegraphics[width=0.35\textwidth]{contour-legend}\\
  \vspace{-0.8em}
  \subfloat[Mahalanobis distance]{\includegraphics[width=0.45\textwidth, height=0.12\textwidth]{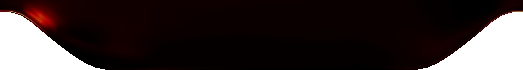}}\hspace{0.1em}
  \subfloat[KDE distance]{\includegraphics[width=0.45\textwidth, height=0.12\textwidth]{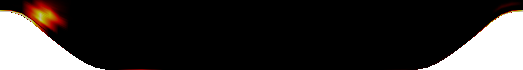}}\\
  \caption{Prediction of Reynolds stress anisotropy for the flow over periodic hills at $Re=10595$ along (a)
    horizontal direction $\xi$ and (b) vertical direction $\eta$ of Barycentric triangle. \textbf{The training set is the flow in a wavy channel at \boldmath{$Re=360$}}. The profiles are shown at 10 locations, $x/H$ = 0,
    0.5, 1, 2, ..., 8. The benchmark discrepancy is obtained based on the Reynolds stress from LES
    simulation~\cite{breuer2009flow}. The contours of (c) Mahalanobis distance and (d) KDE distance are presented for comparison.}
\label{fig:bary_WC}
\end{figure}

After demonstrating the positive correlation between prediction error and extrapolation metrics through an integral view as shown in Fig.~\ref{fig:MSE}, the next step is to investigate the relationship between local prediction performance and extrapolation metrics as shown in Figs.~\ref{fig:bary_Re5600} to~\ref{fig:bary_WC}. 

The predictions of Reynolds stress anisotropy as shown in
Figs.~\ref{fig:bary_Re5600}(a) and~\ref{fig:bary_Re5600}(b) have a good agreement with the benchmark data, which
demonstrates that the prediction performance is satisfactory. The training set is the flow over periodic hills at
$Re=5600$. It should be noted that such satisfactory prediction performance is not
  necessarily guaranteed for Reynolds number extrapolation. In this work, there is no significant
  change of flow physics for the flows over periodic hills from $Re=5600$ to $Re=10595$, which explains the successful Reynolds number extrapolation as shown in Fig.~\ref{fig:bary_Re5600}. For other flows such as boundary layer transition, it is unlikely to achieve the similar quality of prediction by training on the boundary layer without transition and predicting  one with a transition. Therefore, the satisfactory Reynolds number extrapolation results shown here should be interpreted with caution. The KDE distance is close to zero in most areas as shown in Fig.~\ref{fig:bary_Re5600}(d), indicating that the extrapolation extent from training set to test set is small. It is consistent with the prediction performance of Reynolds stress anisotropy as shown in Figs.~\ref{fig:bary_Re5600}(a) and~\ref{fig:bary_Re5600}(b). In addition, it can be seen in Fig.~\ref{fig:bary_Re5600}(d) that the KDE distance is large near the bottom wall at the regions from $x/H=1$ to $x/H=2$ and from $x/H=2$ to $x/H=3$. The reason is that the training set only has data along $x/H=1$, 2 and 3, and the extrapolation extent is expected to be high between these lines due to the flow separation, which leads to a rapid change of flow features. Compared to KDE distance, the Mahalanobis distance is generally high near the bottom wall at the region from $x/H=1$ to $x/H=3$, which indicates that Mahalanobis distance is a more rough estimation of extrapolation compared with KDE distance.

The prediction of Reynolds stress anisotropy as shown in Fig.~\ref{fig:bary_CBFS} is less satisfactory than the prediction as shown in Fig.~\ref{fig:bary_Re5600}, when the flow over curved backward facing step at $Re=13200$ is used as the training set. Specifically, the
prediction of anisotropy shown in Figs.~\ref{fig:bary_CBFS}(a) and~\ref{fig:bary_CBFS}(b)
still have a good agreement with the benchmark data at the regions from $x/H=2$ to $x/H=4$ (the mean squared error of $\xi$ and $\eta$ are 0.0046 and 0.0095, respectively). The prediction of anisotropy is less satisfactory at the downstream region from $x/H=5$ to $x/H=7$ (the mean squared error of $\xi$ and $\eta$ are 0.0124 and 0.0097, respectively), and the prediction is even less satisfactory near the hill crest at inlet and within the contraction region from $x/H=7$ to $x/H=9$ (the mean squared error of $\xi$ and $\eta$ are 0.036 and 0.037, respectively). A
possible reason is that the mean flow feature of contraction and the favorable pressure gradient is
not covered in the training set, which means that the extrapolation extent at the contraction region is
greater. Due to the periodic inlet boundary condition, it is expected that the extrapolation extent near the inlet would also be high. From the contour of KDE distance as shown in Fig.~\ref{fig:bary_CBFS}(d), it can be
seen that such a pattern of extrapolation extent is faithfully represented. Compared to the KDE distance, the Mahalanobis distance as shown in Fig.~\ref{fig:bary_CBFS}(c) is less informative. It demonstrates that the normalized Mahalanobis distance is less able to estimate the extrapolation extent based on geometry extrapolation.

Compared to the prediction of anisotropy as shown in Figs.~\ref{fig:bary_Re5600} and~\ref{fig:bary_CBFS}, the prediction performance is much less satisfactory as shown in Fig.~\ref{fig:bary_WC}, where the training case is the flow in a wavy channel at $Re=360$. The main reason is that the Reynolds number of the training set is much smaller than that of the prediction set, and extrapolation extent is expected to be high across the whole domain. From the contours of Mahalanobis distance and KDE distance as shown in Figs.~\ref{fig:bary_WC}(c) and~\ref{fig:bary_WC}(d), it can be
seen that such a pattern of extrapolation extent is represented by both the
Mahalanobis distance and KDE distance. Both metrics are near one for almost the entire extent of the domain.

% Main point5: The Mahalanobis distance and the prediction performance shows positive correlation
% for the data from a specific case. However, the prediction error can be different for different
% cases with similar Mahalanobis distance. It is because that the increasing rate of prediction
% error may not be the same for different directions in the high dimensional space. However, the
% Mahalanobis distance is only a measure of distance and carries no information of the direction in
% high dimensional space.

%Main point 6: Fig.~\ref{fig:MSE} shows that the magnitude level of prediction error can be
% estimated based on Mahalanobis distance. It indicates that the increasing rate of prediction error
% is not sensitive to a small change of direction in high dimensional space. However, the result
% with C-D channel as training set demonstrates that the increasing rate can be significantly higher
% as shown in Fig.~\ref{fig:Mdist-err} for some specific directions in high dimensional space. The
% more detailed analysis requires the measure of direction in high dimensional space and will be
% investigated in the future work.

\begin{figure}[!htbp]
  \centering
  \subfloat[Training set: PH5600]{\includegraphics[width=0.495\textwidth]{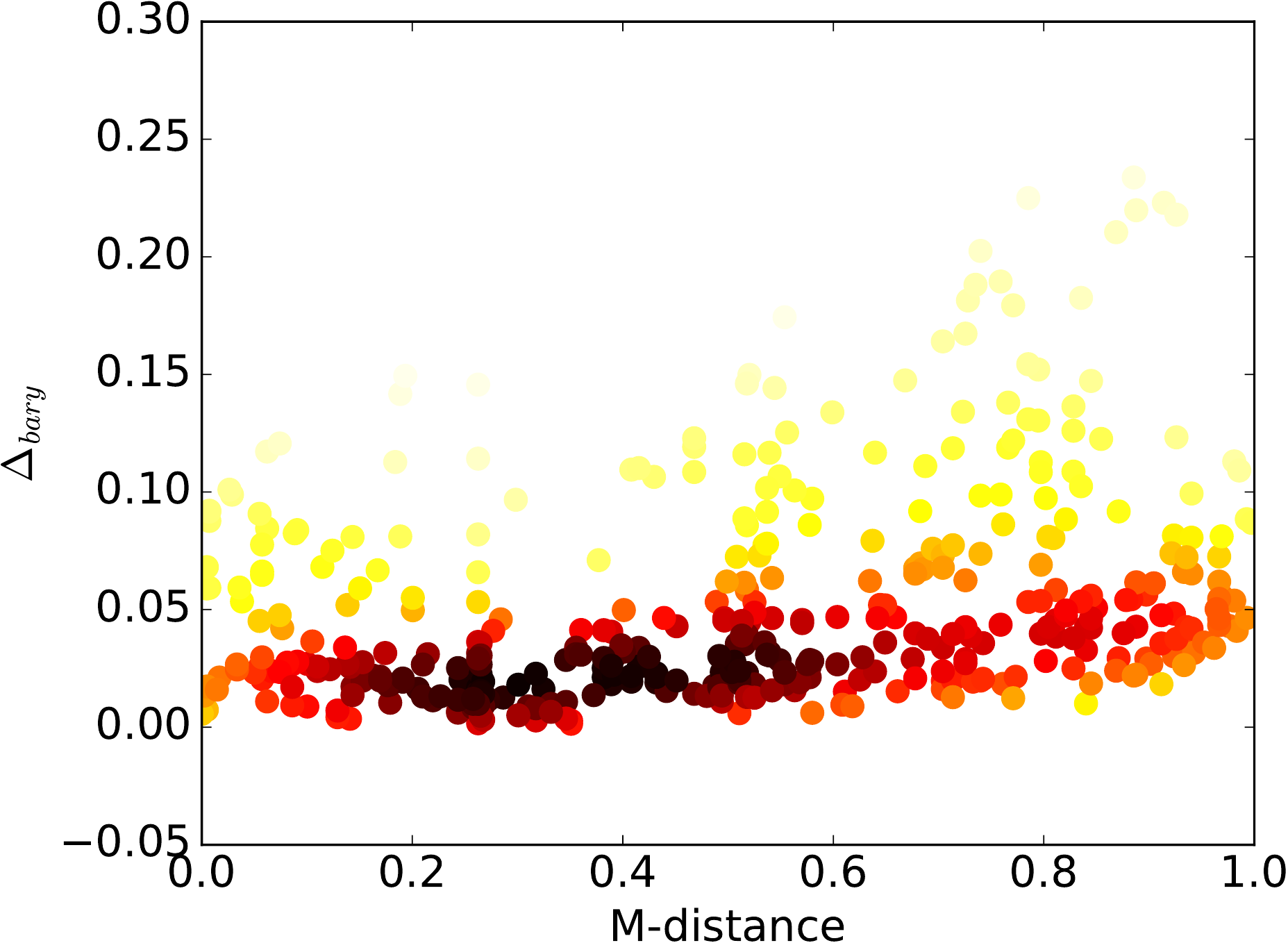}}\hspace{0.1em}
  \subfloat[Training set: PH1400]{\includegraphics[width=0.495\textwidth]{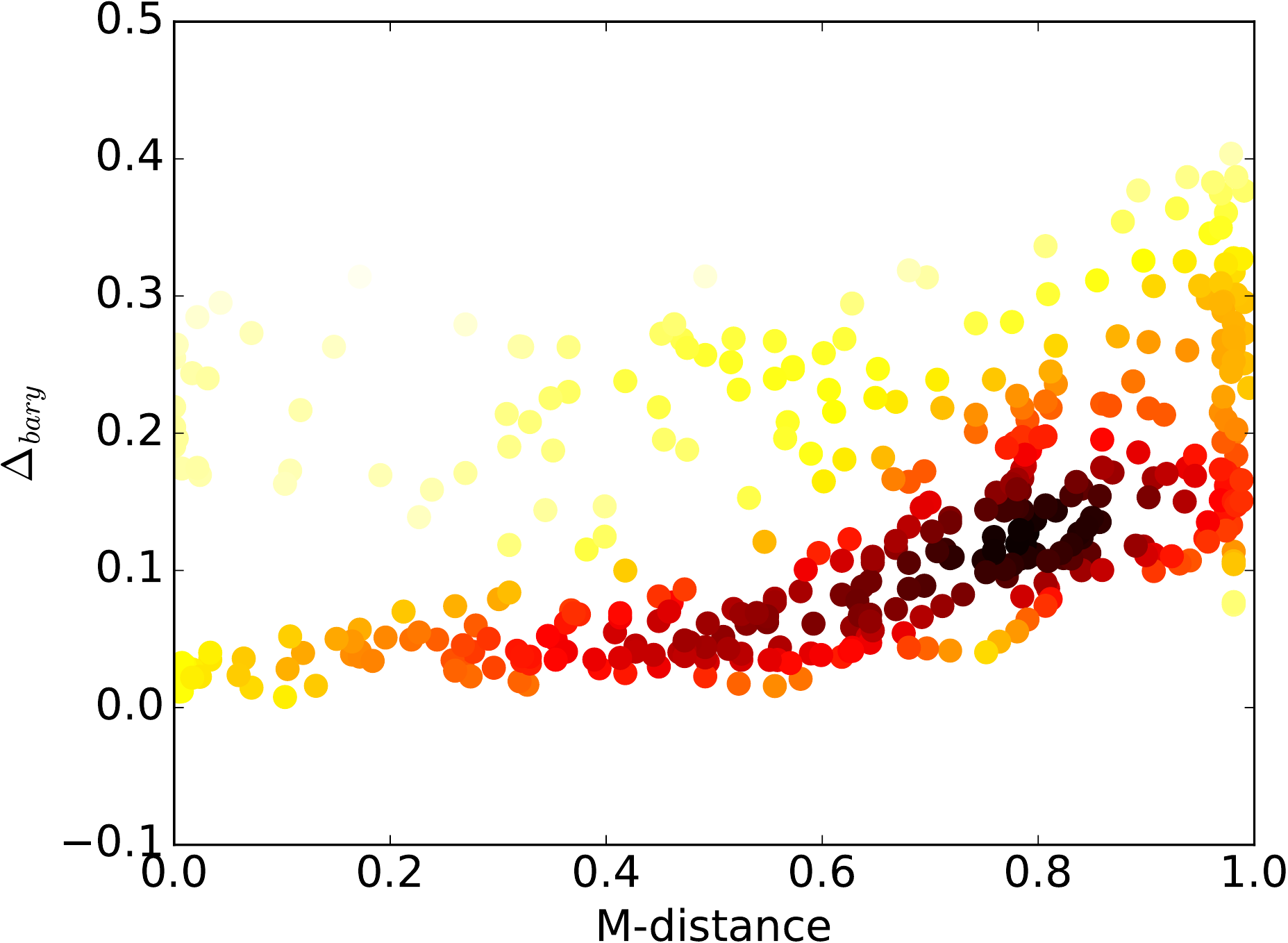}}\\
  \subfloat[Training set: CBFS13200]{\includegraphics[width=0.495\textwidth]{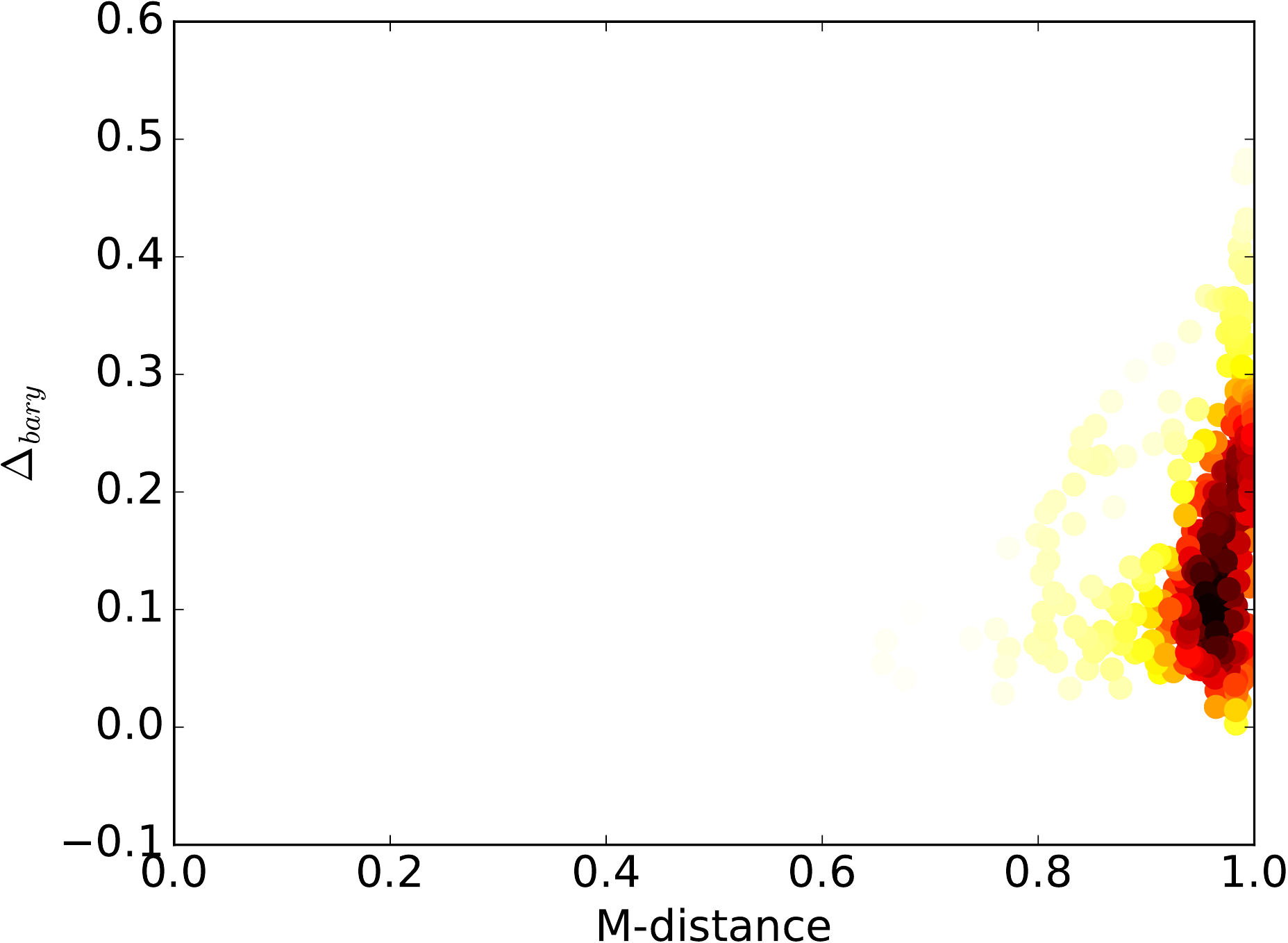}}\hspace{0.1em}
  \subfloat[Training set: WC360]{\includegraphics[width=0.495\textwidth]{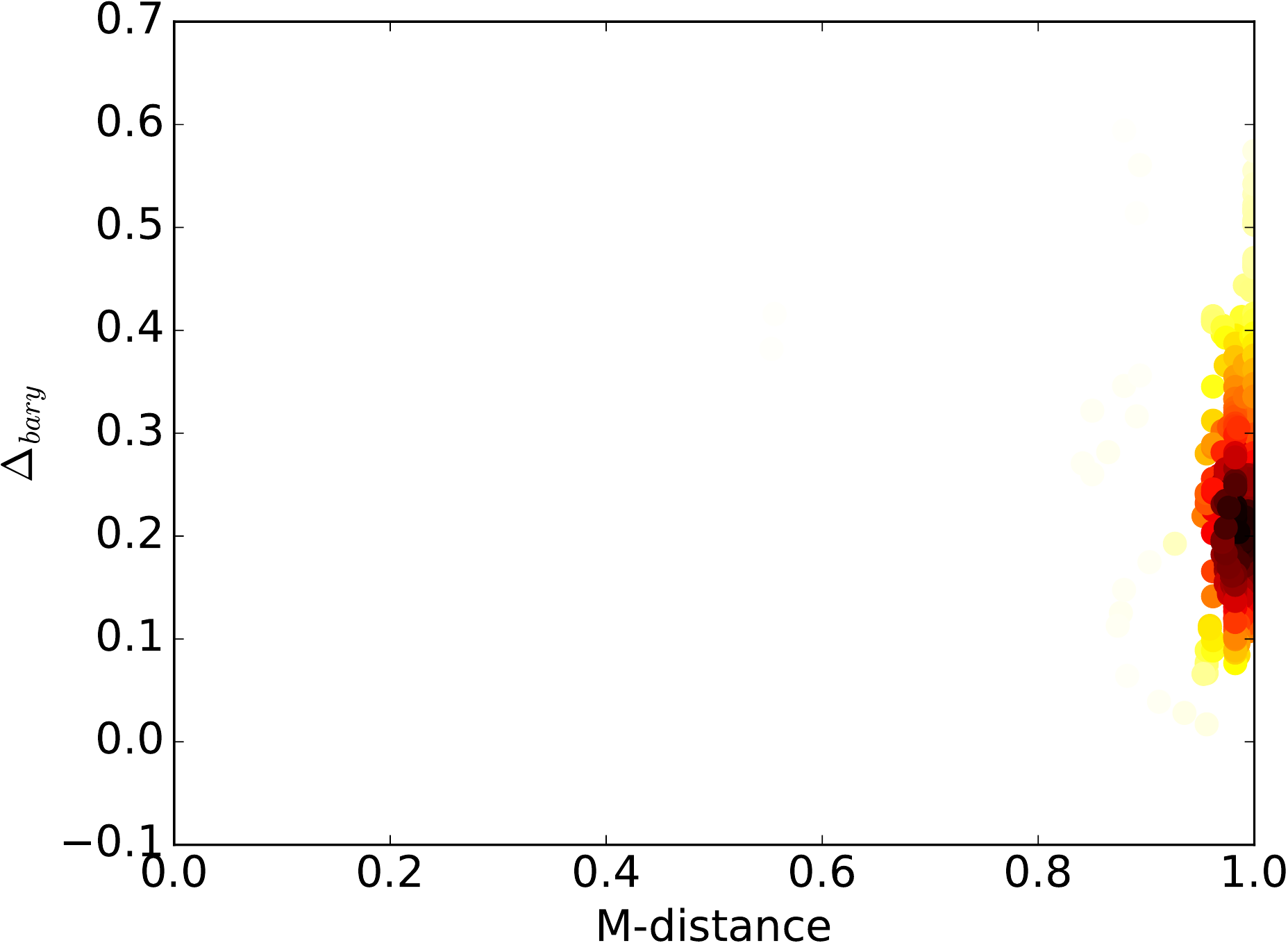}}
  \caption{The correlation between the Mahalanobis distance and the prediction error of Reynolds stress anisotropy. The results based on four training sets are presented, including (a) the flow over periodic hills at $Re=5600$ (PH5600), (b) the flow over periodic hills at $Re=1400$ (PH1400), (c) the flow over curved backward facing step at $Re=13200$ (CBFS13200) and (d) the flow in a wavy channel at $Re=360$ (WC360). Points are colored by the local density of the scatter plot, and the brighter color indicates the higher density.}
\label{fig:Mdist-err}
\end{figure}

\begin{figure}[!htbp]
  \centering
  \subfloat[Training set: PH5600]{\includegraphics[width=0.495\textwidth]{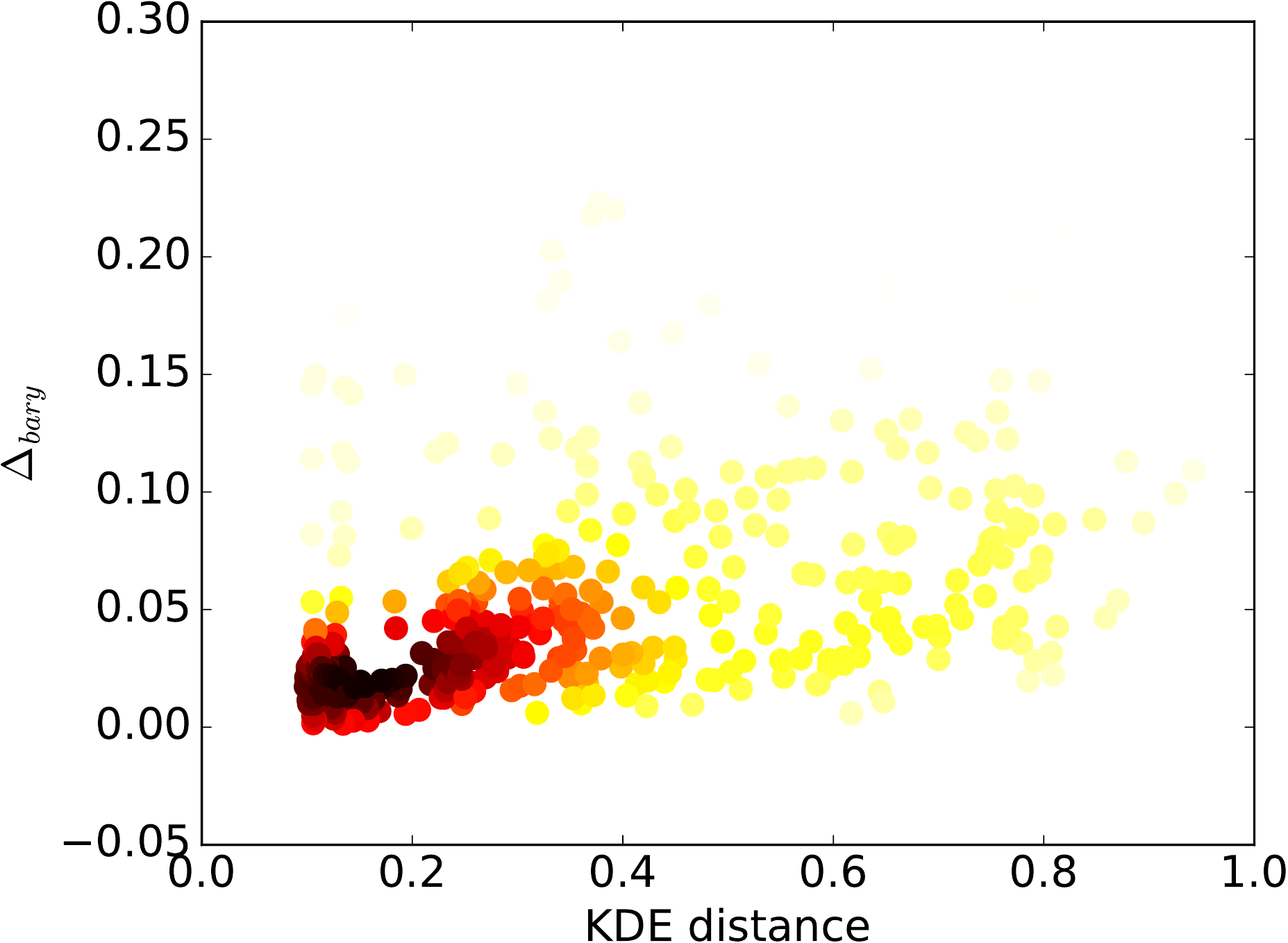}}\hspace{0.1em}
  \subfloat[Training set: PH1400]{\includegraphics[width=0.495\textwidth]{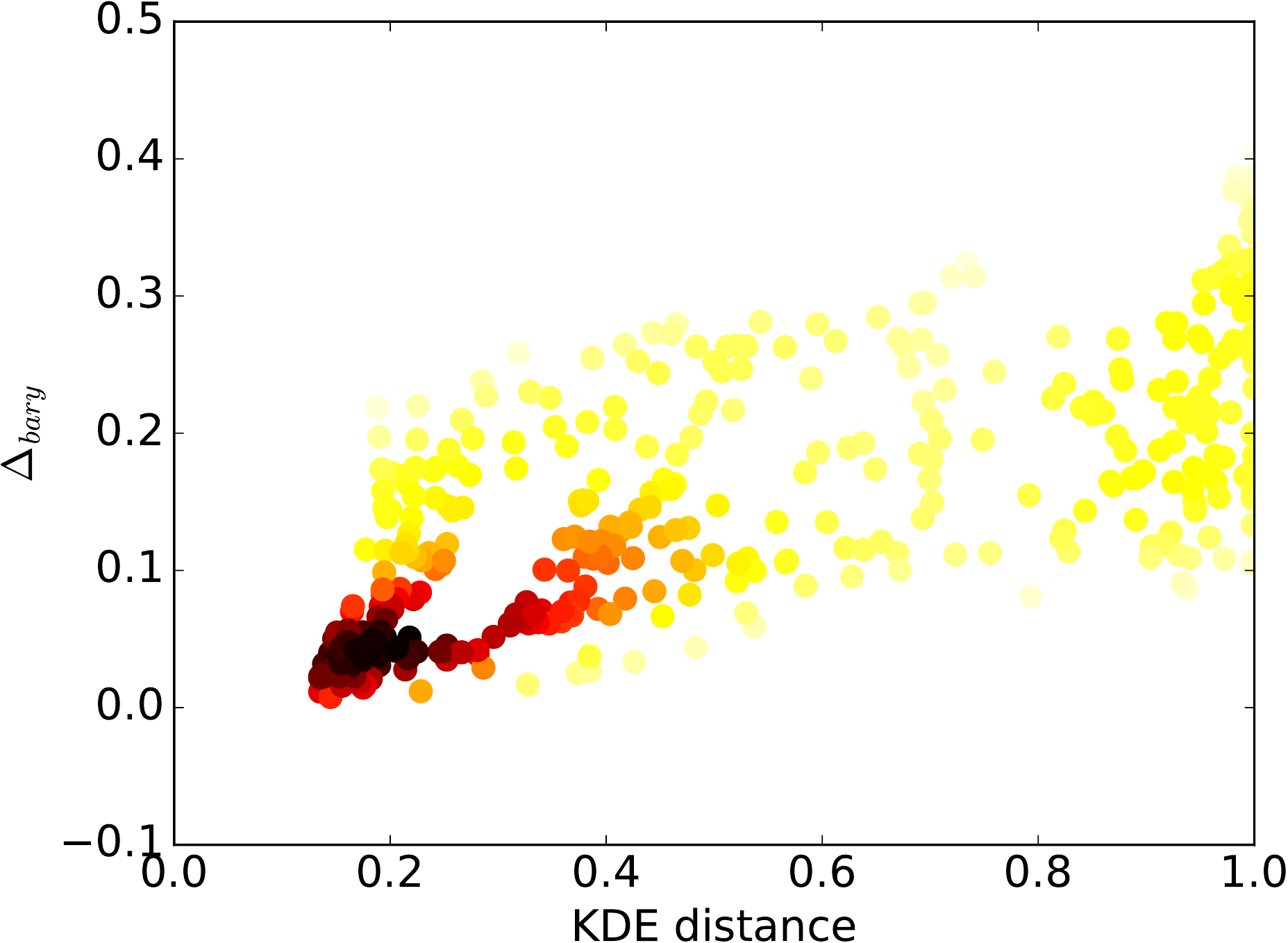}}\\
  \subfloat[Training set: CBFS13200]{\includegraphics[width=0.495\textwidth]{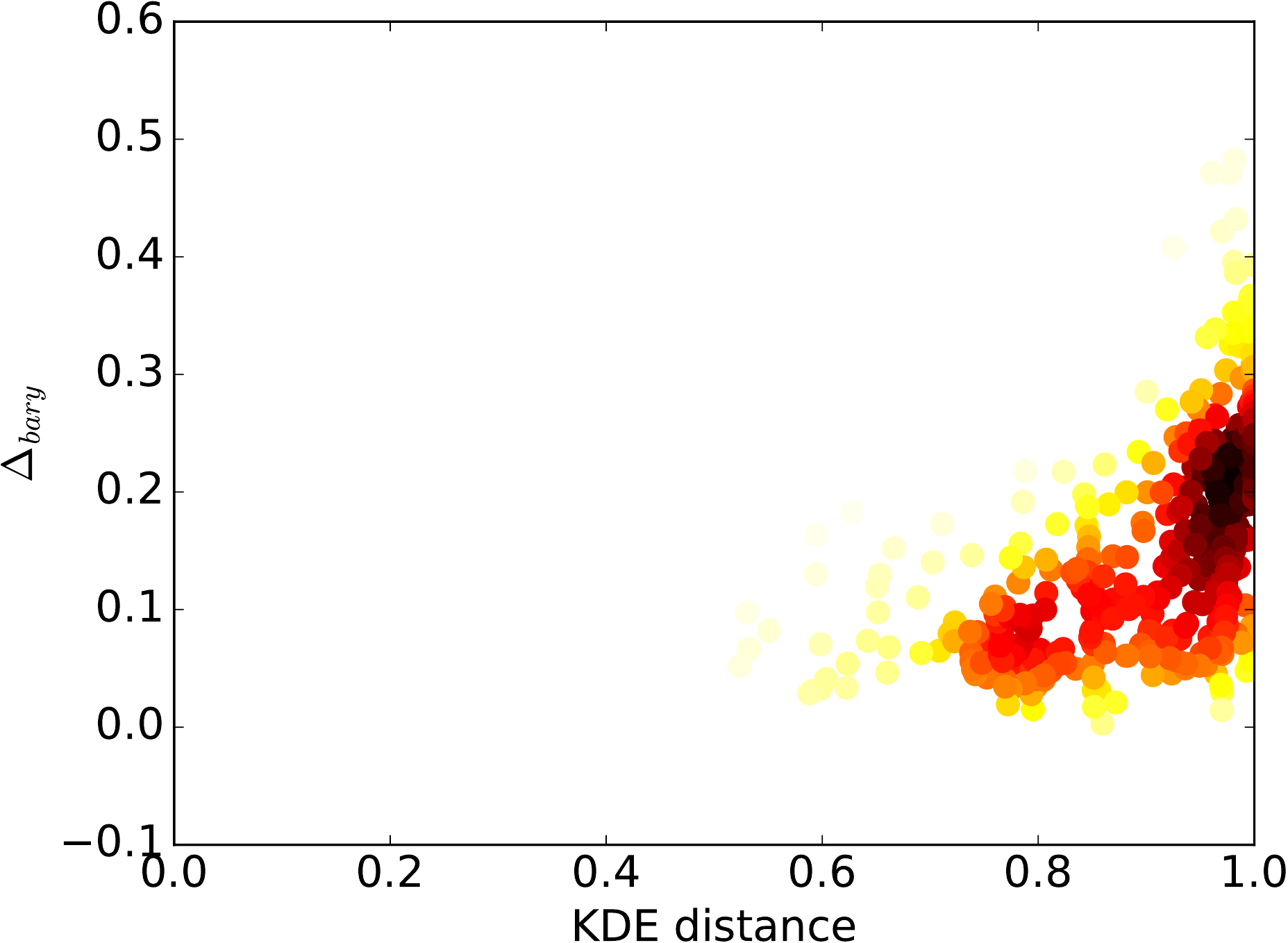}}\hspace{0.1em}
  \subfloat[Training set: WC360]{\includegraphics[width=0.495\textwidth]{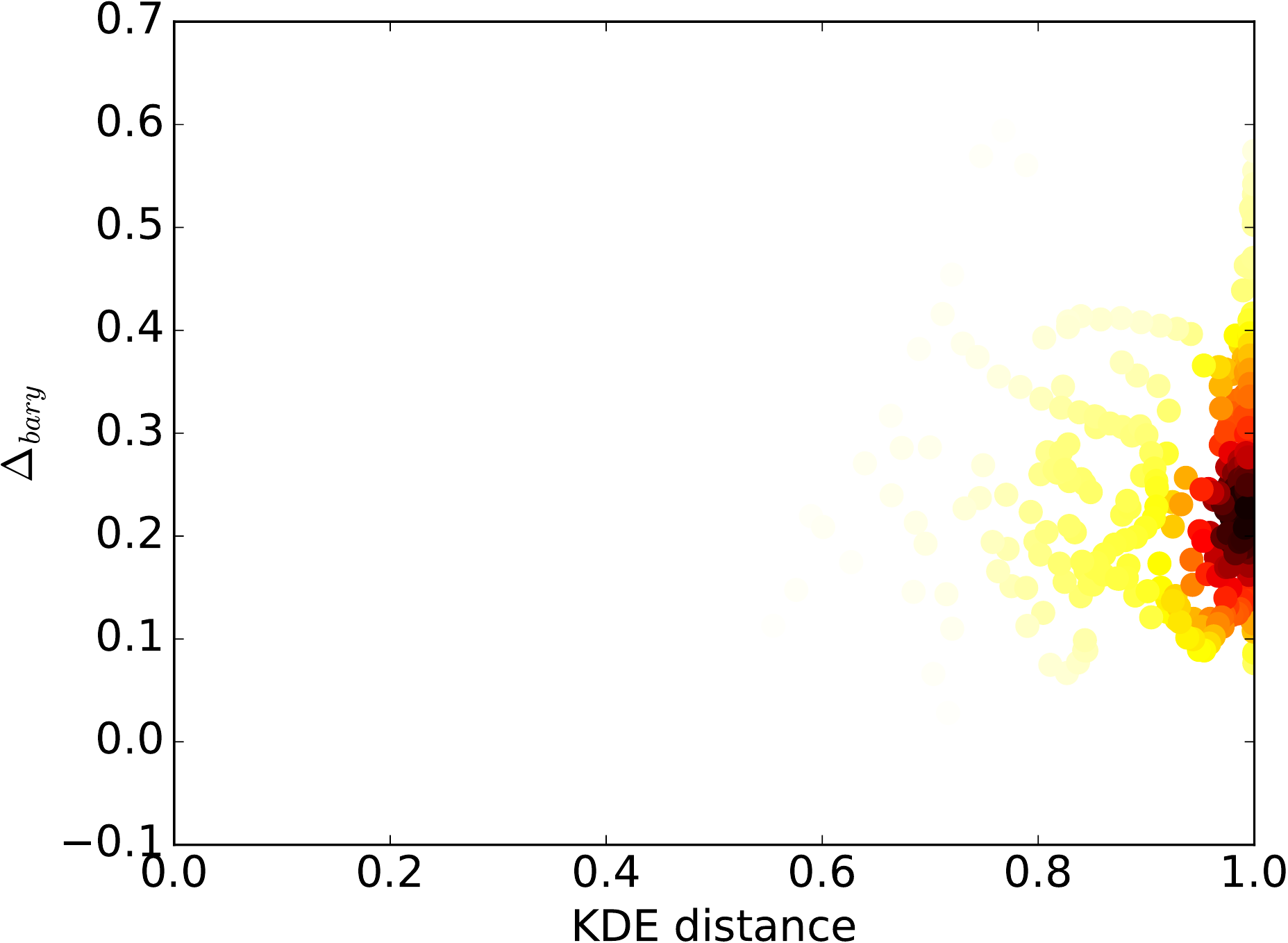}}
  \caption{The correlation between the KDE distance and the prediction error of Reynolds stress anisotropy.  The results based on four training sets are presented, including (a) the flow over periodic hills at $Re=5600$ (PH5600), (b) the flow over periodic hills at $Re=1400$ (PH1400), (c) the flow over curved backward facing step at $Re=13200$ (CBFS13200) and (d) the flow in a wavy channel at $Re=360$ (WC360). Points are colored by the local density of the scatter plot, and the brighter color indicates the higher density.}
\label{fig:kde-err}
\end{figure}

The scatter plot of local prediction error of Reynolds stress anisotropy and extrapolation metrics are presented in Figs.~\ref{fig:Mdist-err} and~\ref{fig:kde-err}. The correlation coefficients are also presented in Table.~\ref{tab:corr} for a more quantitative comparison. 

The scatter pattern is more random as shown in Fig.~\ref{fig:Mdist-err}(a) when the Mahalanobis distance is employed as extrapolation metric and the training set is the flow over periodic hills at $Re=5600$. A possible explanation is that when Mahalanobis distance becomes small, it is expected that the extrapolation error is also small. Therefore, other error sources, such as non-local effect that is not correlated with Mahalanobis distance, will dominate and the relationship between prediction error and Mahalanobis distance tends to become more random.

\begin{table}[!ht]
\begin{center}
\caption {Correlation coefficients between extrapolation metrics and the prediction error of anisotropy}
\label{tab:corr}
\begin{tabular}{P{6cm}P{4cm}P{3cm}}
\toprule
Training Set & Mahalanobis Distance & KDE Distance \\ \midrule
Periodic Hills ($Re=5600$) & 0.31 & 0.40 \\
Periodic Hills ($Re=2800$) & 0.38 & 0.66 \\
Periodic Hills ($Re=1400$) & 0.40 & 0.70 \\
Curved Step ($Re=13200$) & 0.28 & 0.53 \\
C-D Channel ($Re=11300$) & 0.42 & 0.54 \\
Backward Step ($Re=4900$) & 0.41 & 0.57 \\
Wavy Channel ($Re=360$) & $-0.06$ & 0.08 \\
\bottomrule
\end{tabular}
\end{center}
\end{table}

Compared to the correlation shown in Fig.~\ref{fig:Mdist-err}(a), it can be seen in
Figs.~\ref{fig:Mdist-err}(b) and~\ref{fig:Mdist-err}(c) that the correlation is stronger when the training set is the
flow over periodic hills at $Re=1400$ or the flow over curved backward facing step, which is also confirmed by the quantitative comparison of correlation coefficients as shown in Table~\ref{tab:corr}. In these cases, therefore, it can be seen that the extrapolation error is a dominant source of error in the regression models. If
the extrapolation extent further increases, it is shown in Fig.~\ref{fig:Mdist-err}(d) that the
correlation between Mahalanobis distance and prediction error becomes weak again. This is due to the clustering of Mahalanobis distance near one, which significantly reduces the resolution of Mahalanobis distance. In such a scenario, the correlation between prediction error of Reynolds stress anisotropy and Mahalanobis distance is weaker, which leads to a more random relationship between Mahalanobis distance and prediction error.

The pattern of correlation between the prediction error of Reynolds stress anisotropy and KDE distance as shown in Fig.~\ref{fig:kde-err} is similar to that based on Mahalanobis distance. Specifically, there exists a positive correlation if the training set is the flow over periodic hills at $Re=1400$ or the flow over curved backward step at $Re=13200$. The correlation is weaker if the training set is the flow over periodic hills at $Re=5600$ or the flow in a wavy channel at $Re=360$. Although the correlation pattern is similar based on Mahalanobis distance and KDE distance, the clustering pattern is different. The Mahalanobis distance based on Reynolds number extrapolation is more evenly distributed from zero to one, while the KDE distance based on Reynolds number extrapolation tends to cluster near zero. On the other hand, the Mahalanobis distance based on geometry extrapolation is more likely to cluster near one as shown in Fig.~\ref{fig:Mdist-err}(d), while such clustering is less noticeable for KDE distance as shown in Fig.~\ref{fig:kde-err}(d).

The correlation coefficient is smaller based on Mahalanobis distance as shown in Table~\ref{tab:corr}, indicating that the prediction error of Reynolds stress anisotropy is less correlated with Mahalanobis distance. In addition, the correlation coefficient based on Mahalanobis distance can become negative if the extrapolation extent is high, e.g., the flow in a wavy channel at $Re=360$ is used as training set. Therefore, Mahalanobis distance is less accurate in estimating the extrapolation extent and prediction performance compared with the KDE distance. This is consistent with our expectation since the Gaussian distribution assumption in calculating Mahalanobis distance may not be appropriate. However, it should be noted that the Gaussian distribution assumption reduces the memory usage and the computational cost of extrapolation metrics, which is the advantage of the Mahalanobis distance.

\section{Conclusion}
\label{sec:conclusion}
Recently, the increasing interest in data-driven turbulence modeling has created a demand for simple metrics for evaluating extrapolation. Such \textit{a priori} evaluation can provide an estimate of the predictive performance in real applications where high fidelity data are not available for the flow to be predicted. In addition, it can guide the choice of training flows to achieve better prediction performance. In the present work we discuss the evaluation of extrapolation
between different flows in feature space. The
Mahalanobis distance and KDE distance are used as two extrapolation metrics in feature
space to measure the closeness between different flows. Specifically, the flow over periodic hills at $Re=10595$ is used as the test set in this work. Three training sets at different Reynolds numbers and four training sets with
different geometries are individually used to investigate the relationship between the prediction error and the extrapolation metrics. In particular, two training sets at different Reynolds numbers and two training sets with different geometries
are chosen for detailed analysis. The results demonstrate that the relationship between extrapolation metrics and the prediction error is less correlated in two extreme scenarios, i.e., when the training set is very similar to the test set or very different from the test set. In the former case, other sources of error besides extrapolation error dominate the prediction uncertainty. In the later case, the degree of extrapolation is so high that the machine learning algorithm is just ``guessing'', leading to a plateau in the error rate. Except for these two extreme scenarios, both the Mahalanobis distance and the KDE distance are positively correlated with the prediction error, demonstrating that both extrapolation metrics can be used in estimating the extrapolation extent and prediction performance. However, the quantitative comparison of correlation coefficient shows that the prediction error is less correlated with the Mahalanobis distance, indicating that the estimation of extrapolation extent based on Mahalanobis distance is less accurate. On the other hand, the Gaussian distribution assumption of Mahalanobis distance reduces the memory usage and the computational cost, which is the advantage of using Mahalanobis distance. In conclusion, the KDE distance is preferable if the accuracy of extrapolation estimation is more important, while the Mahalanobis distance is still acceptable in some applications in which the memory usage or the computational cost of high dimensional kernel density estimation is not affordable.

\section*{Acknowledgment}

HX would like to thank Dr. Eric G. Paterson for numerous helpful discussions during this work.

Sandia National Laboratories is a multi-program laboratory managed and operated by Sandia
Corporation, a wholly owned subsidiary of Lockheed Martin Corporation, for the U.S. Department of
Energy's National Nuclear Security Administration under contract DE-AC04-94AL85000. SAND2016-6700J

\section*{Compliance with Ethical Standards}
Conflict of Interest: The authors declare that they have no conflict of interest.

%% References


%\appendix

% \input{materials-xiao.tex}

\bibliographystyle{elsarticle-num}
%\bibliography{learning-confidence}

\begin{thebibliography}{10}
\expandafter\ifx\csname url\endcsname\relax
  \def\url#1{\texttt{#1}}\fi
\expandafter\ifx\csname urlprefix\endcsname\relax\def\urlprefix{URL }\fi
\expandafter\ifx\csname href\endcsname\relax
  \def\href#1#2{#2} \def\path#1{#1}\fi

\bibitem{Craft}
T.~Craft, B.~Launder, K.~Suga, Development and application of a cubic
  eddy-viscosity model of turbulence, International Journal of Heat and Fluid
  Flow 17 (1996) 108--115.

\bibitem{Milano}
M.~Milano, P.~Koumoutsakos, Neural network modeling for near wall turbulent
  flow, Journal of Computational Physics 182 (2002) 1--26.

\bibitem{Tracey2013}
B.~Tracey, K.~Duraisamy, J.~Alonso, Application of supervised learning to
  quantify uncertainties in turbulence and combustion modeling, AIAA Aerospace
  Sciences Meeting, AIAA 2013-0259, 2013.

\bibitem{Duraisamy2015}
K.~Duraisamy, Z.~Shang, A.~Singh, New approaches in turbulence and transition
  modeling using data-driven techniques, AIAA 2015--1284, 2015.

\bibitem{ling2015evaluation}
J.~Ling, J.~Templeton, Evaluation of machine learning algorithms for prediction
  of regions of high reynolds averaged navier stokes uncertainty, Physics of
  Fluids (1994-present) 27~(8) (2015) 085103.

\bibitem{Ling2016}
J.~Ling, A.~Ruiz, G.~Lacaze, J.~Oefelein, Uncertainty analysis and data-driven
  model advances for a jet-in-crossflow, {ASME Turbo Expo 2016}.

\bibitem{ling2016reynolds}
J.~Ling, A.~Kurzawski, J.~Templeton, Reynolds averaged turbulence modelling
  using deep neural networks with embedded invariance, Journal of Fluid
  Mechanics 807 (2016) 155--166.

\bibitem{Wang2016}
J.-X. Wang, J.-L. Wu, H.~Xiao, A physics informed machine learning approach for reconstructing 
Reynolds stress modeling discrepancies based on DNS data, Physical Review Fluids 2(3), 034603 (2017).

\bibitem{ling2016jcp}
J.~Ling, R.~Jones, J.~Templeton, Machine learning strategies for systems with
  invariance properties, Journal of Computational Physics 318 (2016) 22--35.

\bibitem{liaw2002classification}
A.~Liaw, M.~Wiener, Classification and regression by randomforest, R news 2~(3)
  (2002) 18--22.

\bibitem{gorle2012rans}
C.~Gorle, M.~Emory, G.~Iaccarino, {RANS} modeling of turbulent mixing for a jet
  in supersonic cross flow: model evaluation and uncertainty quantification,
  in: ICHMT DIGITAL LIBRARY ONLINE, Begel House Inc., 2012.

\bibitem{emory2011modeling}
M.~Emory, R.~Pecnik, G.~Iaccarino, Modeling structural uncertainties in
  reynolds-averaged computations of shock/boundary layer interactions, AIAA
  paper 479 (2011) 2011.

\bibitem{xiao-mfu}
H.~Xiao, J.-L. Wu, J.-X. Wang, R.~Sun, C.~J. Roy, Quantifying and reducing
  model-form uncertainties in {Reynolds-Averaged} {Navier-Stokes} simulations:
  A data-driven, physics-based {Bayesian} approach, Journal of Computational
  Physics 324 (2016) 115--136.

\bibitem{banerjee2007presentation}
S.~Banerjee, R.~Krahl, F.~Durst, C.~Zenger, Presentation of anisotropy
  properties of turbulence, invariants versus eigenvalue approaches, Journal of
  Turbulence 8~(32) (2007) 1--27.

\bibitem{breiman2001random}
L.~Breiman, Random forests, Machine learning 45~(1) (2001) 5--32.

\bibitem{friedman2001elements}
J.~Friedman, T.~Hastie, R.~Tibshirani, The elements of statistical learning,
  Springer, Berlin, 2001.

\bibitem{emory2013modeling}
M.~Emory, J.~Larsson, G.~Iaccarino, Modeling of structural uncertainties in
  {Reynolds}-averaged {Navier}-{Stokes} closures, Physics of Fluids 25~(11)
  (2013) 110822.

\bibitem{nasa-web}
G.~H. Chris~Rumsey, Brian~Smith,
  \href{http://turbmodels.larc.nasa.gov}{Turbulence modeling resource} (2016).
\newline\urlprefix\url{http://turbmodels.larc.nasa.gov}

\bibitem{breuer2009flow}
M.~Breuer, N.~Peller, C.~Rapp, M.~Manhart, Flow over periodic hills--numerical
  and experimental study in a wide range of reynolds numbers, Computers \&
  Fluids 38~(2) (2009) 433--457.

\bibitem{bentaleb2012large}
Y.~Bentaleb, S.~Lardeau, M.~A. Leschziner, Large-eddy simulation of turbulent
  boundary layer separation from a rounded step, Journal of Turbulence~(13)
  (2012) N4.

\bibitem{laval2011direct}
J.-P. Laval, M.~Marquillie, Direct numerical simulations of
  converging--diverging channel flow, in: Progress in Wall Turbulence:
  Understanding and Modeling, Springer, 2011, pp. 203--209.

\bibitem{le1997direct}
H.~Le, P.~Moin, J.~Kim, Direct numerical simulation of turbulent flow over a
  backward-facing step, Journal of fluid mechanics 330 (1997) 349--374.

\bibitem{maass1996direct}
C.~Maa{\ss}, U.~Schumann, Direct numerical simulation of separated turbulent
  flow over a wavy boundary, in: Flow Simulation with High-Performance
  Computers II, Springer, 1996, pp. 227--241.

\bibitem{Cover1967}
T.~Cover, P.~Hart, Nearest neighbor pattern classification, {IEEE Transactions}
  13 (1967) 21--27.

\bibitem{mahalanobis1936generalized}
P.~C. Mahalanobis, On the generalized distance in statistics, Proceedings of
  the National Institute of Sciences (Calcutta) 2 (1936) 49--55.

\bibitem{Silverman}
B.~Silverman, Density estimation for statistics and data analysis, {CRC Press},
  1986.

\bibitem{scott2015multivariate}
D.~W. Scott, Multivariate density estimation: theory, practice, and
  visualization, John Wiley \& Sons, 2015.

\bibitem{launder1974application}
B.~Launder, B.~Sharma, Application of the energy-dissipation model of
  turbulence to the calculation of flow near a spinning disc, Letters in heat
  and mass transfer 1~(2) (1974) 131--137.

\bibitem{weller1998tensorial}
H.~G. Weller, G.~Tabor, H.~Jasak, C.~Fureby, A tensorial approach to
  computational continuum mechanics usi ng ject-oriented techniques, Computers
  in physics 12~(6) (1998) 620--631.

\bibitem{thompson2016methodology}
R.~L. Thompson, L.~E.~B. Sampaio, F.~A. de~Bragan{\c{c}}a~Alves, L.~Thais,
  G.~Mompean, A methodology to evaluate statistical errors in {DNS} data of
  plane channel flows, Computers \& Fluids 130 (2016) 1--7.

\bibitem{poroseva2016accuracy}
S.~V. Poroseva, J.~D. Colmenares~F, S.~M. Murman, On the accuracy of {RANS}
  simulations with {DNS} data, Physics of Fluids 28~(11) (2016) 115102.

\bibitem{Wang2017}
J.-X. Wang, J.-L. Wu, J.~Ling, G.~Iaccarino, H.~Xiao, A comprehensive
  physics-informed machine learning framework for predictive turbulence
  modeling, submitted. Available at arXiv:1701.07102 (2017).

\end{thebibliography}

%\input{notation-paper}
\end{document}